\documentclass[11pt]{article}

\usepackage[labelfont=bf,labelsep=period]{caption}
\usepackage{realA4}

\usepackage{float}
\usepackage[grumpy]{gitinfo}
\usepackage{graphicx}
\usepackage{times}
\usepackage{url}
\usepackage{xspace}
\usepackage[detect-all,binary-units]{siunitx}
\newcommand{\appr}{{\mathord{\sim}}}
\usepackage{tikz}
\usetikzlibrary{calc,shapes,arrows,matrix,positioning,decorations.pathreplacing}
\usepackage{subfigure}

\usepackage{booktabs}

\usepackage{amsmath}
\allowdisplaybreaks[2]

\usepackage{amssymb}

\floatstyle{ruled}
\newfloat{algo}{htb}{alg}
\floatname{algo}{Algorithm}

\newcommand{\N}{\mathbb{N}}

\newcommand{\false}{\str{false}\xspace}
\newcommand{\true}{\str{true}\xspace}

\usepackage[nodayofweek]{datetime}
\newdateformat{simple}{\THEDAY\ \monthname[\THEMONTH]\ \THEYEAR}
\simple

\newcommand{\str}[1]{\textsc{#1}}

\newcommand{\op}[1]{\textsl{#1}}
\newcommand{\msg}[2]{\ensuremath{[\str{#1}, {#2}]}}
\newcommand{\concat}{\ensuremath{\circ}}

\newcommand{\tup}[1]{%
  \relax\ifmmode
    \langle #1 \rangle%
  \else
    $\langle$#1$\rangle$%
  \fi}

\newcommand{\CA}{\ensuremath{\mathcal{A}}\xspace}
\newcommand{\CK}{\ensuremath{\mathcal{K}}\xspace}
\newcommand{\CO}{\ensuremath{\mathcal{O}}\xspace}
\newcommand{\CQ}{\ensuremath{\mathcal{Q}}\xspace}
\newcommand{\CR}{\ensuremath{\mathcal{R}}\xspace}
\newcommand{\CS}{\ensuremath{\mathcal{S}}\xspace}
\newcommand{\CU}{\ensuremath{\mathcal{U}}\xspace}
\newcommand{\CV}{\ensuremath{\mathcal{V}}\xspace}
\newcommand{\CZ}{\ensuremath{\mathcal{Z}}\xspace}

\newcommand{\nil}{\str{null}}

\newcommand{\ASSERT}{\textbf{assert}\xspace}

\newcommand{\DO}{\textbf{do}\xspace}
\newcommand{\ELSE}{\textbf{else}\xspace}
\newcommand{\FOR}{\textbf{for}\xspace}
\newcommand{\FUNCTION}{\textbf{function}\xspace}

\newcommand{\IF}{\textbf{if}\xspace}

\newcommand{\RETURN}{\textbf{return}\xspace}
\newcommand{\STATE}{\textbf{state}\xspace}

\newcommand{\THEN}{\textbf{then}\xspace}
\newcommand{\UPON}{\textbf{upon}\xspace}

\newcommand{\shortlist}{\parskip0pt\topsep0pt\partopsep0pt\parsep1pt\itemsep2pt}

\newcommand{\NAME}{VICOS\xspace}
\newcommand{\forklin}{fork-lin\-e\-ar\-iza\-bi\-lity\xspace}

\usepackage{listings}
\definecolor{backcolor}{rgb}{0.95,0.95,0.95}

\lstdefinestyle{logstyle}{
    backgroundcolor=\color{backcolor},   
    basicstyle=\footnotesize,
    frame=single,
    breakatwhitespace=false,                                         
}

\lstdefinestyle{xmlstyle}{
    language=XML,
    basicstyle=\small,
    breakatwhitespace=false,         
    keepspaces=true,
    frame=single,                                  
}

\begin{document}

\title{\bf Don't Trust the Cloud, Verify: 
  Integrity and Consistency for Cloud Object Stores\footnote{A short
  version of this work appears in the proceedings of SYSTOR 2015. 
  DOI: 10.1145/2757667.2757681}}

\author{%
Marcus Brandenburger\\
IBM Research - Zurich\\
\url{bur@zurich.ibm.com}
\and
Christian Cachin\\
IBM Research - Zurich\\
\url{cca@zurich.ibm.com}
\and
Nikola Kne\v{z}evi\'{c}\footnote{Work done at IBM Research~-Zurich}\\
IMC Financial Markets\\
\url{nikkne@gmx.ch}
}

\date{\gitCommitterDate}

\maketitle

\begin{abstract}
  \noindent
  Cloud services have turned remote computation into a commodity and enable
  convenient online collaboration.  However, they require that clients
  fully trust the service provider in terms of confidentiality, integrity,
  and availability.  Towards reducing this dependency, this paper
  introduces a protocol for \emph{verification of integrity and consistency
    for cloud object storage (\NAME)}, which enables a group of mutually
  trusting clients to detect data-integrity and consistency violations for
  a cloud object-storage service.  It aims at services where multiple
  clients cooperate on data stored remotely on a potentially misbehaving
  service.  \NAME enforces the consistency notion of \forklin, supports
  wait-free client semantics for most operations, and reduces the
  computation and communication overhead compared to previous protocols.
  \NAME is based in a generic way on any \emph{authenticated data
    structure}.  Moreover, its operations cover the hierarchical name space
  of a cloud object store, supporting a real-world interface and not only a
  simplistic abstraction.  A prototype of \NAME that works with the
  key-value store interface of commodity cloud storage services has been
  implemented, and an evaluation demonstrates its advantage compared to
  existing systems.
\end{abstract}

\section{Introduction}
\label{sec:introduction}

More and more data is outsourced to the cloud, and
collaborating on a \emph{shared resource} using cloud services has
become easier than ever.  Programmers work together on online source
code repositories, global project teams
produce technical deliverables, and friends share their photo albums.
Nevertheless, the clients need to \emph{trust} the cloud provider as
rely on it for the confidentiality and correctness of their data.
\emph{Encryption} may preserve the confidentiality of data but cannot
prevent inadvertent or malicious data modifications.  This work shows
how to protect the \emph{integrity} and \emph{consistency} of data
on an untrusted cloud storage service accessed by multiple clients.

With a single client only, the client may locally keep a short
\emph{cryptographic hash value} of the outsourced data.  Later, this
can be used to verify the integrity of the data returned by the cloud
storage service. However, with multiple disconnected
clients, no common synchronization, and no communication among the
clients, neither hashing nor digital signatures are sufficient by
themselves.  The reason is that a malicious or \emph{Byzantine}
server may violate the consistency of the data, for example, by
reordering or omitting properly authenticated operations, so that the
\emph{views} of the storage state at different clients diverge.  A
malicious cloud server may pretend to one set of clients that some
operations by others simply did not occur.  In other words,
\emph{freshness} can be violated and the clients cannot detect such
replay attacks until they communicate directly.  The problem is particularly
relevant in cryptographic online voting and for web certificate
transparency~\cite{laurie14}.

The strongest achievable notion of consistency in this multi-client
model is captured by \emph{\forklin}, introduced by Mazi{\`e}res and
Shasha~\cite{mazsha02}.  A consistency and integrity verification
protocol may guarantee this notion by adding condensed data about the
causal evolution of the client's views into their interaction with the
server.  This ensures that if the server creates only a single
discrepancy between the views of two clients, these clients may never
observe operations of each other afterwards.  In other words, if the
server ever lies to some clients and these clients communicate later,
they will immediately discover the violation.  Hence, with only one
check they can verify a large number of past operations.

The SUNDR system~\cite{lkms04} pioneered fork-linearizable consistency
and demonstrated a network file system protected by a hash tree of every
user's files.  SUNDR uses an expensive protocol, requiring messages of
size~$\Omega(n^2)$ for $n$ clients~\cite{cashsh07}.  Like other systems
providing \forklin, it suffers from the inherent limitation that
sometimes, even with a correct server, clients have to block and cannot
proceed with their next operation because other clients are concurrently
executing operations~\cite{cashsh07}.

In order to prevent blocking, FAUST~\cite{cakesh11}, SPORC~\cite{fzff10},
and Venus~\cite{scckms10} relax their guarantees to \emph{weak \forklin},
which establishes consistency only eventually, after further operations
occur; this is not desirable because a client may only know later that a
protocol output was not correct.  SPORC and the related Blind Stone Tablet
(BST)~\cite{wisish09} protocol shift the maintenance of state to the
clients, such that the server is merely responsible for coordination; every
client holds a complete copy of the system's state.  This contradicts the
goal of outsourcing data to the cloud.  Another way to avoid blocking is
explored by BST and COP~\cite{cacohr14}: they let any commuting operations
proceed immediately.

In this paper, we present \emph{\NAME}, a \emph{verification protocol for
  the integrity and consistency of cloud object storage}, which overcomes
these limitations and demonstrates a complete practical system.  \NAME
supports the optimal consistency notion of \forklin, provides wait-free
semantics for all \emph{compatible} client operations, and has smaller
overhead than previous protocols.  The notion of compatible operations, as
introduced here, generalizes the progress condition over commuting
operations, which are considered in past work.  Informally, two operations
$o$ and $\omega$ are compatible if executing $o$ before $\omega$ does not
influence the behavior of $\omega$ (neither its effects nor its output
value).
Conceptually, \NAME is based on abstract \emph{authenticated data
  structures} in a modular way.  Moreover, it unifies the two different
lines of work on this problem so far, namely, the untrusted storage
protocols~\cite{lkms04,cashsh07,cakesh11} that feature remote state and are
based on vector clocks, and the remote service verification
protocols~\cite{wisish09,fzff10,cacohr14}, which create local copies of the
state and use hash chains.  In particular, \NAME maintains state remotely,
at the server, but uses hash chains for consistency verification.
Furthermore, \NAME incurs only a constant communication overhead per
operation, independent of the number of clients, whereas FAUST and Venus
require vector clocks of size~$\Omega(n)$ with $n$~clients.

We have implemented and evaluated \NAME with a commodity cloud object
store; the results demonstrate that the overhead for distributed,
multi-client integrity and consistency verification is low.  \NAME protects
a complete cloud storage service (assuming it has atomic semantics),
spanning many objects and offering read, write, delete, and directory
listing operations; this stands in contrast to Venus~\cite{scckms10}, which
only provided consistency for a \emph{single} data object.  Furthermore,
\NAME notifies the clients whenever the integrity or consistency is
violated but does not address recovery operations.

\subsection{Contributions}

This work makes the following contributions towards ensuring integrity
for data stored on untrusted cloud providers:
\begin{itemize}
\item A novel abstract protocol to verify the integrity and consistency
  of a generic service based on an authenticated data
  structure~\cite{tamass03}, which ensures \forklin, supports wait-free
  semantics for compatible operations, and incurs only constant
  communication overhead.  This protocol also generalizes authenticated 
  data structures to multiple writers.

\item An instantiation of this protocol for a commodity cloud object
  store, called \NAME.  It represents the first integrity protection protocol
  with all of the above features for the standard operations of a cloud
  object store.

\item An implementation and evaluation of \NAME using \emph{COSBench},
  demonstrating its practicality.  In particular, the overhead of \NAME
  for integrity protection remains acceptable small with moderate
  concurrency and increases slightly when many clients access the same
  data concurrently.  The prototype is available open-source.
\end{itemize}

\subsection{Organization}

The paper continues with introducing the model and defines our notion of
an authenticated data structure (ADS).  Sec.~\ref{sec:generic-protocol}
presents the abstract integrity protocol for ADS and discusses its
properties.  \NAME is introduced in Sec.~\ref{sec:frieda} and the
evaluation appears in Sec.~\ref{sec:evaluation}.  Related work is
discussed in Sec.~\ref{sec:related_work}.

\section{Model}
\label{sec:model}

We consider an asynchronous distributed system with~$n$ mutually
trusting \emph{clients}~$C_1,...,C_n$ and a \emph{server}~$S$.
The communication between the server and the clients is reliable and
respects \emph{first-in first-out (FIFO)} semantics.  Clients cannot
communicate with each other.
A protocol~$P$ specifies the behavior of the clients and the server. All
clients are \emph{correct} and hence follow~$P$; in particular, they do
not crash (although crashes could be tolerated with extra measures in the 
protocols).
The server is either correct and follows~$P$ or \emph{Byzantine},
deviating arbitrarily from~$P$.

The clients invoke operations of a stateful functionality~$F$,
implementing a set of deterministic operations; $F$ defines a response
and a state change for every operation.  We use the standard notions of
executions, histories, sequential histories, real-time order,
concurrency, and well-formed executions from the
literature~\cite{AttiyaW04}.  In particular, every operation in an
execution is represented by an \emph{invocation event} and a
\emph{response event}.  We extend $F$ with a special return value
\str{abort} that allows an operation to abort without taking effect,
which may be used when concurrent operations would cause it to
block~\cite{mdss09,cacohr14}.

\subsection{Consistency properties}
\label{sec:defconsistency}

When $S$ is correct, the protocols should provide the standard notion of
\emph{linearizability}~\cite{herwin90} with respect to~$F$.  It requires
that an execution of operations by all clients together is equivalent to an
imaginary sequential execution of $F$.  More precisely, for an execution to
be linearizable, all invocation and response events occuring at all clients
can be permuted into one totally ordered sequence, where (1) every
operation invocation is followed immediately by the corresponding response;
(2) all operations are correct according to the specification of $F$; and
(3) the operations in the sequence respect the real-time order among
operations observed in the execution.

Fork-linearizability~\cite{mazsha02, cashsh07} relaxes this common global
view and permits that the clients observe an execution that may split into
multiple linearizable ``forks,'' which must never join again.  More
precisely, an execution is \emph{fork-linearizable} when every client
observes a linearizable history (containing all operations that the client
executes itself) and for any operation observed by multiple clients, the
history of events occurring before the operation is the same at those
clients.  This implies that if the views of the execution at two clients
ever diverge, they cannot observe each other's operations any more, and
makes it easy to spot consistency violations by the server.

Furthermore, we recall the concept of a \emph{fork-linearizable
Byzantine emulation}~\cite{cashsh07}.  It requires that our protocol
among the clients and the Byzantine server satisfies two conditions:
When the server is correct, then the service is linearizable; otherwise,
it is still fork-linearizable.  Finally, our protocol may only abort (by
returning \str{abort}) if there is some reason for this; in other words,
when the clients execute operations sequentially, then no client ever
aborts.

\subsection{Cryptographic primitives}
\label{sec:defcrypto}

Our protocols use a cryptographic \emph{hash function} and
\emph{digital signature schemes} for protecting data against
modification.  We model them in an idealized way, as if implemented by a
distributed oracle~\cite{cachin2011introduction}.

A cryptographic \emph{hash function} \op{hash} maps a bit string~$x$ of
arbitrary length to a short, unique hash value~$h$.  Its implementation
is deterministic and maintains a list $L$ of all $x$ that have been
queried so far.  When the invocation contains $x\in L$, then \op{hash}
responds with the index of $x$ in $L$; otherwise, \op{hash} appends $x$
to $L$ and returns its index.  This ideal implementation models only
collision resistance, i.e., that it is not feasible to find two
different values~$x_1$ and $x_2$ such that~$\op{hash}(x_1) = \op{hash}(x_2)$.

A \emph{digital signature scheme} as used here provides two functions,
$\op{sign}_i$ and $\op{verify}_i$, to ensure the authenticity of a
message created by a known client.  The scheme works as follows: A
client~$C_i$ invokes~$\op{sign}_i(m)$ with a message~$m$ as argument and
obtains a signature~$\phi \in \{0,1\}^*$ with the response.  Only
client~$C_i$ can invoke~$\op{sign}_i$.  When $m$ and $\phi$ are sent to
another client, that client may verify the integrity of~$m$.  It
invokes~$\op{verify}_i(\phi, m)$ and obtains \true as a response if and
only if $C_i$ has executed $\op{sign}_i(m)$;
otherwise,~$\op{verify}_i(\phi,m)$ returns~\false.  Every client, as
well as~$S$, may invoke \op{verify}.

\subsection{Authenticated data structures}
\label{sec:defads}

This section defines the model of \emph{authenticated data structures}
(ADS) used here.
Authenticated data structures~\cite{naonis00,mndgks04,tamass03} are a
well-known tool for verifying operations and their results over data
outsourced to untrusted servers. Popular instantiations rely on Merkle
hash trees or other hierarchical authenticated search
structures~\cite{gotatr11,mhks14}.

We model an ADS for an arbitrary deterministic \emph{functionality}~$F$.
Departing from the literature on ADSs, we eliminate the special role of
the single writer or ``source'' and let any client perform update
operations; likewise, we unify queries and updates into one type of
\emph{operation} from a set~\CO.  Operations may contain arguments
according to~$F$, but these are subsumed into the different~$o \in \CO$.
The functionality specifies a state~$s \in \CS$, which will be
maintained by the server, starting with an initial state~$s_{0}$.  For
example, this may include all data stored on a cloud storage service.
Given $s$, applying an operation $o$ of $F$ means to compute~$(s', r)
\gets F(s, o)$, resulting in a new state~$s' \in \CS$ and a response~$r
\in \CR$.

Operations are executed by the clients and formally described by an
invocation event and a response event occurring at the client.  In order
to verify the responses of~$S$, a client stores an
\emph{authenticator}~$a$, a short value also called a \emph{digest}.
Initially, the authenticator is a special value~$a_0$.

For executing an operation~$o$, the client invokes
algorithm~$\op{query}_F$ on~$S$,
\[
  (r, \sigma_o) \ \gets \ \op{query}_F(s, o),
\]
producing a \emph{response}~$r$ and \emph{auxiliary data}~$\sigma_o$
for~$o$; the latter may serve as a \emph{proof} for the validity of the
response.  Then the client locally performs an operation
$\op{authexec}_F$ to validate the response on the basis of the authenticator.
From this the client obtains an output
\[
  (a', \sigma'_o, v) \ \gets \ \op{authexec}_F(o, a, r, \sigma_o).
\]
Here $a'$ and $\sigma'_0$ are the updated authenticator and auxiliary
data, respectively, and $v \in \{\false, \true\}$ denotes a Boolean
verification value that tells the client whether the response~$r$
from~$S$ is valid.

The client should then send $\sigma'_o$ back to~$S$, so that the server
may actually execute~$o$ and update its state from $s$ to $s'$ by
running $\op{refresh}_F$ as
\[
  s' \ \gets \ \op{refresh}_F(s, o, \sigma'_o).
\]
Note we may also consider these operations for \emph{sequences} of
operations.

An ADS~\cite{naonis00,mndgks04,tamass03} is a special case of this
formalization, in which the operations \CO can be partitioned into
\emph{update-operations}~\CU and \emph{query-operations}~\CQ.  Update
operations generate no response, i.e., $F(s, u) = (s', \bot)$ for all $u
\in \CU$ and queries do not change the state, that is, $F(s, q) = (s,
r)$ for all $q \in \CQ$.

Furthermore, an ADS may contain initialization and key-generation
routines and all algorithms may take public and private keys as inputs
in addition.  For simplicity, and because our ADS implementations are
unkeyed, we omit them here.

An ADS satisfies \emph{correctness} and \emph{security}.  Consider the
execution of a sequence $\langle o_1, \dots, o_m \rangle$ of operations,
$F(s_0, \langle o_1, \dots, o_m \rangle)$, which means to compute $(s_j,
r_j) \gets F(s_{j-1}, o_j)$ for $j=1, \dots, m$.  A \emph{proper
authenticated execution} of $\langle o_1, \dots, o_m \rangle$ computes
the steps
\begin{align*}
  (r_j, \sigma_{o,j}) \ &\gets \ \op{query}_F(s_{j-1}, o_j) \\
  (a_{j}, \sigma'_{o,j}, v_j) 
               \ &\gets \ \op{authexec}_F(o_j, a_{j-1}, r_j, \sigma_{o,j}) \\
  s_j          \ &\gets \ \op{refresh}_F(s_{j-1}, o_j, \sigma'_{o,j}).
\end{align*}
such that $v_j = \true$, for $j=1, \dots, m$.

An ADS is \emph{correct} if the proper authenticated execution of any
operation sequence $\langle o_1, \dots, o_m \rangle$ outputs state~$s_m$
and response~$r_m$ such that $(s_m, r_m) = F(s_0, \langle o_1, \dots,
o_m \rangle)$.

Furthermore, an ADS must be \emph{secure} against an adversary~\CA that
tries to forge a response and auxiliary data that are considered valid
by a client.  More precisely, \CA adaptively determines an operation
sequence $\langle o_1, \dots, o_m \rangle$, which is taken through a
proper authenticated execution by a challenger; at every step, \CA
obtains $a_j$ and $\sigma'_{o,j}$ and then determines $o_{j+1}$ and
whether the execution continues.  Finally, after obtaining $a_m$ and
$s_m$, \CA outputs an operation $o^*$, a response~$r^*$, and a value
$\sigma_o^*$.  The ADS is secure if no \CA succeeds in creating $o^*$,
$r^*$, $s_m^*$, and $\sigma_o^*$ such that
\[
  (\cdot, \cdot, \true) \ = \ \op{authexec}_F(o^*, a_m, r^*, \sigma_o^*)
\]
but $F(s_m, o^*) = (s_{m+1}, r_{m+1})$ with $r_{m+1} \neq r^*$.  
(The ``don't-care'' symbol $\cdot$ in the tuple indicates that
only a subset of the tuple elements are needed.)
In other words, \CA cannot find any $o^*$
executed on $s_m$ and forge a response~$r^*$ and a proof for $o^*$ that is
accepted by the client, unless the response is correct according to~$F$ and
$r^* = r_{m+1}$.

Note that this formalization represents an ``idealized'' security
notion.  It is easy to formulate an equivalent computational security
condition using the language of modern cryptology~\cite{Goldreich01}.
This model subsumes the one of Cachin~\cite{cachin11} and generalizes
ADS~\cite{tamass03} to multiple writing clients, where here the
authenticator is implicitly synchronized among the clients.

\subsection{Compatible operations}
\label{sec:defcompatible}

Our protocol takes advantage of \emph{compatible} operations that permit
``concurrent'' execution without compromising the goal of ensuring
\forklin to the clients.  An operation~$o'$ is compatible with another
operation $o$ (in a state~$s$) if the presence of $o'$ before $o$ does
not influence the return value of~$o$ (in~$s$).  Compatible operations
can be executed without blocking; this improves the throughput compared
to earlier protocols, in particular with respect to COP~\cite{cacohr14},
which considered the stronger notion of commutative operations.

Formally, we say an operation sequence~$\omega$ is
\emph{compatible with} an operation~$o$ \emph{in a state~$s$} 
whenever the responses of $o$ remain
the same regardless of whether~$\omega$ executed before~$o$. Hence, with
\begin{align*}
  (s', r)  \ &\gets \ F(s, \omega); \quad
  (s'', p) \  \gets \ F(s', o);     \quad \text{and} \\
  (t', q)  \ &\gets \ F(s, o)
\end{align*}
it holds $p = q$.
Moreover, we say that $\omega$ is \emph{compatible} with~$o$ if and only
if $\omega$ is compatible with $o$ in all states $s \in \CS$ of~$F$.

In algorithms we use a function~$\op{compatible}_F$, which takes
$\omega$ and $o$ as inputs and returns \true if and only if $\omega$ and
$o$ are compatible.  

In the terminology of the database literature~\cite{WeikumV02}, two
operations are compatible if and only if they do not exhibit a
``write-read'' conflict, which is also known as a ``dirty read.''

Note that compatible operations are not necessarily commutative, but
commutative operations are always compatible.
For instance, a query operation~$q$ is compatible with an update~$u$ in
any state; but when $q$ returns data modified by~$u$, then $q$ and $u$
do not commute.

\section{The ADS integrity protocol (AIP)}
\label{sec:generic-protocol}

This section introduces the \emph{ADS integrity protocol (AIP)}, a generic
protocol to verify the integrity and consistency for any authenticated data
structure (ADS) operated by a remote untrusted server.
AIP extends and improves upon the \emph{commutative-operation verification
  protocol (COP)} and its authenticated variant (ACOP) of Cachin and
Ohrimenko~\cite{cacohr14}.  Sec.~\ref{sec:frieda} shows how to instantiate
AIP with an authenticated dictionary for protecting cloud storage; the
result forms the core of \NAME.

\subsection{Overview}
\label{sec:overview}

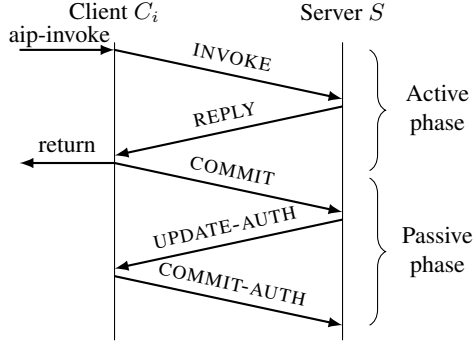
\begin{figure}
    \centering
        \begin{tikzpicture}[-latex, font=\footnotesize]

            \foreach \i in {0,1} {
                \node (X-\i) at (3*\i,0) {};
            }

            \foreach \i in {0,...,6} {
                \node (Y-\i) at (0,0.75*\i) {};
            }

            \draw[-] ($(X-0) + (Y-0) + (0,-0.1)$) -- ($(X-0) + (Y-5) + (0,0.1)$);
            \draw[-] ($(X-1) + (Y-0) + (0,-0.1)$) -- ($(X-1) + (Y-5) + (0,0.1)$);
            
            \node[] at ($(X-0) + (Y-5) + (0,0.5)$) {Client~$C_i$};            
            \node[] at ($(X-1) + (Y-5) + (0,0.5)$) {Server~$S$};

            \draw[thick] ($(X-0) + (Y-5)$) -- ($(X-1) + (Y-4) + (0,0.1)$) node[midway,sloped,above] {\str{invoke}};
            \draw[thick] ($(X-1) + (Y-4)$) -- ($(X-0) + (Y-3) + (0,0.1)$) node[midway,sloped,above] {\str{reply}};
            \draw[thick] ($(X-0) + (Y-3)$) -- ($(X-1) + (Y-2) + (0,0.1)$) node[midway,sloped,above] {\str{commit}};
            
            \draw[thick] ($(X-1) + (Y-2)$) -- ($(X-0) + (Y-1) + (0,0.1)$) node[midway,sloped,above] {\str{update-auth}};
            \draw[thick] ($(X-0) + (Y-1)$) -- ($(X-1) + (Y-0) + (0,0.1)$) node[midway,sloped,above] {\str{commit-auth}};

            \draw[thick] ($(X-0) + (Y-5) + (-1.25,0)$) -- ($(X-0) + (Y-5)$) node[midway,sloped,above, yshift=-0.1cm, xshift=-0.1cm] {aip-invoke};
            \draw[thick] ($(X-0) + (Y-3)$) -- ($(X-0) + (Y-3) + (-1.25,0)$) node[midway,sloped,above] {return};

            \draw[decorate,decoration={brace,raise=3pt,amplitude=6pt}, -]
            ($(X-1) + (Y-5) + (0.25,0)$) -- ($(X-1) + (Y-3) + (0.25,-0.1)$)
            node[midway,xshift=1cm, text width=1.5cm, text centered] {Active \\phase};

            \draw[decorate,decoration={brace,raise=3pt,amplitude=6pt}, -]
            ($(X-1) + (Y-3) + (0.25,0) + (0,-0.2)$) -- ($(X-1) + (Y-0) + (0.25,0.1)$)
            node[midway,xshift=1cm, text width=1.5cm, text centered] {Passive \\phase};

        \end{tikzpicture}
    \caption{Protocol messages in AIP.}
    \label{fig:phases}
\end{figure}

The processing of one operation in AIP is structured into an
\emph{active} and a \emph{passive} phase, as shown in
Fig.~\ref{fig:phases}.  The active phase begins when the client invokes
an operation and ends when the client completes it and outputs a
response; this takes one message roundtrip between the client and the
server.  Different from past protocols, the client stays further
involved with processing authentication data for this operation during
the passive phase, which is decoupled from the execution of further
operations.

More precisely, when client~$C_i$ invokes an operation~$o \in \CO$, it
sends a signed \str{invoke} message carrying~$o$ to the server~$S$.  The
server assigns a global sequence number ($t$) to~$o$ and responds with a
\str{reply} message containing a list of \emph{pending} operations, the
response, an authenticator, and auxiliary data needed by the client for
verification.  Operations are pending (for~$o$) because they have been
started by other clients and $S$ has ordered them before $o$, but~$S$
has not yet finished processing them.  We distinguish between
\emph{pending-other} operations, which have been invoked by other
clients, and \emph{pending-self} operations, which $C_i$ has executed
before~$o$.

After receiving the \str{reply} message, the client checks its content.  In
particular, if the pending-other operations are compatible with~$o$,
then~$C_i$ verifies the pending-self operations including~$o$ with the help
of the authenticator; if they are correct, $C_i$ proceeds and outputs the
response immediately.  Along the way $C_i$ verifies that all data received
from~$S$ satisfies conditions to ensure \forklin.  An operation that
terminates like this is called \emph{successful}; alternatively, when the
pending-other operations are not compatible with~$o$, then $o$
\emph{aborts}.  In this case, $C_i$ returns the symbol \str{abort}.
In any case, the client subsequently \emph{commits}~$o$ and sends a signed
\str{commit} message to~$S$ (note that also aborted operations are
committed in that sense).  This step terminates the active phase of the
operation.  The client may now invoke the next operation or retry~$o$ if it
was aborted.

Processing of~$o$ continues with the passive phase.  At some (later) time,
as soon as the operation immediately preceding~$o$ in the assigned order
has terminated its own passive phase, $S$ sends an \str{update-auth}
message with auxiliary data and the authenticator of the preceding
operation to~$C_i$.  When $C_i$ receives this, it validates the message
content and verifies the execution of~$o$ unless $o$ had been aborted.
Using the methods of the ADS, the client now computes and signs a new
authenticator that it sends to~$S$ in a \str{commit-auth} message.  We say
that $C_i$ \emph{authenticates}~$o$ at this time.  When $S$ receives this
message, then it \emph{applies}~$o$ by executing it on the state and stores the
corresponding authenticator; this completes the passive phase of~$o$.

Note that the server may receive \str{commit} messages in an order that
differs from the one of the globally assigned sequence numbers due to
asynchrony.  Still, the authentication steps in the passive phases of
the different operations must proceed according to the assigned
operation order.  For this reason, the server maintains a second sequence
number ($b$), which indicates the last authenticated operation that the
server has applied to its state.  Hence, $S$ buffers the incoming
\str{commit} messages and runs the passive phases sequentially in the
assigned order.

For ensuring consistency, every client needs to know about all
operations that the server has executed.  Therefore, when $S$ responds
to the invocation of an operation by~$C_i$, it includes in the
\str{reply} message a summary (including the corresponding signatures)
of all those operations that $C_i$ has missed since it last executed an
operation.  Prior to committing~$o$, the client verifies these
operations and thereby \emph{clears} them.

\subsection{Notation}

The protocol is shown in Alg.~\ref{alg:client}--\ref{alg:server} and
formulated reactively.
The clients and the server are state machines whose actions are triggered
by events such as receiving messages.  An ordered list with elements $e_1,
e_2, \dots, e_k$ is denoted by $E = \langle e_1, e_2, \dots, e_k \rangle$;
the element with index~$j$ may be accessed as~$E[j]$.  We also use maps
that operate as associative arrays and store values under unique keys.  A
value~$v$ is stored in a map~$H$ by assigning it to a key~$k$, denoted
by~$H[k] \gets v$; for non-assigned keys, the map returns~$\bot$.  
The symbol $\|$ denotes the
concatenation of bit strings.  The \textbf{assert} statement, parameterized
by a condition, catches an error and immediately terminates the protocol
when the condition is false.  Clients use this to signal that the server
misbehaved.

\subsection{Data structures}
\label{sec:datastructures}

This section describes the data structures maintained by every client and
by the server.  For simplicity, the pseudo code does not describe garbage
collection, but we note where this is possible.

Every \emph{client}~$C_i$ (Alg.~\ref{alg:client}) stores the sequence number of its last cleared
operation in a variable~$c$.  The \emph{hash chain}~$H$ represents the
condensed view that $C_i$ has of the sequence of all operations.  It is
computed over the sequence of all applied operations and the sequence of
pending operations announced by~$S$.  Formally, $H$ is a map indexed by
operation sequence number; an entry~$H[l]$ is equal to $\op{hash}(H[l-1]
\| o \| l \| j)$ when the $l$-th operation~$o$ is executed by $C_j$,
with~$H[0] = \nil$.
Variable~$Z$ is a map that represents the status (\str{success} or
\str{abort}) of every operation, according to the result of the test
for compatibility.  The client only needs the entries in~$H$ and~$Z$
with indices greater than~$c$ and may garbage-collect older entries.
Finally, $C_i$ uses a variable~$u$ that is set to~$o$ whenever $C_i$ has
invoked operation~$o$ but not yet completed it; otherwise~$u$ is~$\bot$.

The \emph{server} (Alg.~\ref{alg:server}) maintains the sequence number of the most recently
invoked operation in a counter~$t$.  In addition to that, the
counter~$b$ contains the sequence number of the most recently applied
operation and governs the authentication of operations in the passive
phase.
Every invoked operation is stored in a map~$I$ and every committed
operation in a map~$O$; both maps are indexed by sequence number.  The
server only needs the entries in~$I$ with sequence numbers greater
than~$b$.  An entry in~$O$, at a sequence number $b$ or greater, has to
be stored until every client has committed some operation with a higher
sequence number and may be removed later.
Most importantly, the server keeps the state~$s$ of the ADS for~$F$,
which reflects all successful operations up to sequence number~$b$.  In
contrast to COP~\cite{cacohr14} and SPORC~\cite{fzff10}, where every
client maintains a complete copy of the state, here only the server
stores that state.  Moreover, $S$ stores the authenticator for every
operation in a map~$A$ indexed by sequence number.

\subsection{The protocol in detail}
\label{sec:generic-protocol:details}

This section describes the ADS integrity protocol (AIP) as shown in
Alg.~\ref{alg:client}--\ref{alg:server}.  AIP is parameterized by an ADS
and a functionality~$F$ that specifies its operations through
$\op{query}_F$, $\op{authexec}_F$, and $\op{refresh}_F$.  The client
invokes AIP with an ADS-operation~$o$ by calling $\op{aip-invoke}(o)$;
it completes when AIP executes \RETURN at the end of the handler for the
\str{reply} message.  This ends the active phase of AIP, and the passive
phase continues asynchronously in the background.

\subsubsection{Active phase}

When client~$C_i$ invokes an operation~$o$, it computes an
\str{invoke}-signature~$\tau$ over~$o$ and~$i$; this proves to other
clients that $C_i$ has invoked~$o$.  Then $C_i$ stores $o$ in~$u$ and
sends an \str{invoke} message with $o$ and~$\tau$ to the server.

Upon receiving an \str{invoke} message with~$o$, the server increments
the sequence number~$t$, assigns it to~$o$, and assembles the
\str{reply} message for~$C_i$.  First, $S$ stores $o$ and the
accompanying signature in~$I[t]$; the value $t$ is also called
the \emph{position} of~$o$.  The pending operations for~$o$,
assigned to~$\omega$, are found in $I[b+1], \dots, I[t]$, i.e.,
starting with the oldest non-applied operation, and include~$o$.
In order to compute the response and the auxiliary data for~$o$ from
the correct state, the server must then extract the \emph{successful} 
pending-self operations~$\mu$ of $C_i$, using the following method:

\vbox{ 
\small
\begin{tabbing}
xxxx\=xxxx\=xxxx\=xxxx\=xxxx\=xxxx\=xxxx\kill
\FUNCTION $\op{separate-pending}(i, \omega)$ \\
\> $\mu \gets \langle \rangle$; $\gamma \gets \langle \rangle$  \\
\> \FOR $k = 1, \dots, \op{length}(\omega)$ \DO \\
\> \> $(o', \cdot, j) \gets \omega[k]$ \\
\> \> \IF $j = i$ \THEN \\
\> \> \> \IF $k = \op{length}(\omega) \lor 
         \text{status of $o'$ is $\str{success}$}$ \THEN \\
\> \> \> \> $\mu \gets \mu \concat \langle o' \rangle$
            \` // see text how to get status of $o'$ \\
\> \> \ELSE \IF $j \neq i$ \THEN \\
\> \> \> $\gamma \gets \gamma \concat \langle o' \rangle$ \\
\> \RETURN $(\mu, \gamma)$
\end{tabbing}
}

\begin{algo}[p]
\vbox{
\small
\begin{tabbing}
xxxx\=xxxx\=xxxx\=xxxx\=xxxx\=xxxx\=xxxx\kill
\STATE \\
\> $c \in \N_0$: sequence number of last cleared operation, initially~0 \\
\> $H: \N_0 \to \{0,1\}^*$: the hash chain, initially only $H[0] = \bot$ \\
\> $Z: \N_0 \to \CZ$: status map, initially empty  \\
\> $u \in \CO \cup \{\bot\}$: current operation or~$\bot$ if none, initially~$\bot$ \\
\\
\FUNCTION $\op{aip-invoke}(o)$ \\
\> $u \gets o$ \\
\> $\tau \gets \op{sign}_i(\str{invoke} \| o \| i)$ \\
\> send message \msg{invoke}{o, \tau, c} to $S$ \\
\\
\UPON receiving message \msg{reply}{\delta, b, \alpha, \omega, t, r, \sigma_o} from $S$ \DO \\
\> $(a, \psi) \gets \alpha$ \\
\> \op{check-view}($\delta, b, a, \psi$) \\
\> \op{check-pending}($\omega$) \\
\> $(\mu, \gamma) \gets \op{separate-pending}(i, \omega)$ \\
\> $t \gets b + \op{length}(\omega)$ \\
\> \IF $\op{compatible}_F(\gamma, u)$ \THEN \\
\> \> $(\cdot, \cdot, v) \gets \op{authexec}_F(\mu, a, r, \sigma_o)$ \\
\> \> \ASSERT $v$ \\
\> \> $Z[t] \gets \str{success}$ \\
\> \ELSE \\
\> \> $r \gets \str{abort}$ \\
\> \> $Z[t] \gets \str{abort}$ \\
\> $\phi \gets \op{sign}_i(\str{commit} \| t \| u \| i \| Z[t] \| H[t])$ \\
\> send message \msg{commit}{u, t, Z[t], \phi} to $S$ \\
\> $u \gets \bot $ \\
\> \RETURN $r$
    \` // response of operation $\op{aip-invoke}(o)$ \\
\\
\UPON recv.\ msg.\ \msg{update-auth}{o, r, \sigma_o, \phi, q, 
   \delta, \alpha} from $S$ \DO \\
\> \ASSERT $\op{verify}_i(\phi, \str{commit} \| q \| o \| i \| Z[q] \| H[q])$ \\
\> $(o_\delta, \cdot, \cdot, \cdot, j) \gets \delta$ \\
\> $(a, \psi) \gets \alpha$ \\
\> \ASSERT $\op{verify}_j(\psi, \str{auth} \| o_\delta \| q-1 \| H[q-1] \| a)$ \\
\> \IF $Z[q] = \str{success}$ \THEN \\
\> \> $(a', \sigma'_o, v) \gets \op{authexec}_F(o, a, r, \sigma_o)$ \\
\> \> \ASSERT $v$ \\
\> \ELSE \\
\> \> $(a', \sigma'_0) \gets (a, \bot)$ \\
\> $\psi' \gets \op{sign}_i(\str{auth} \| o \| q \| H[q] \| a')$ \\
\> send message \msg{commit-auth}{a', \sigma'_o,\psi'} to $S$
\end{tabbing}
}
\caption{ADS integrity protocol (AIP) for client~$C_i$}
\label{alg:client}
\end{algo}

\begin{algo}
\vbox{
\small
\begin{tabbing}
xxxx\=xxxx\=xxxx\=xxxx\=xxxx\=xxxx\=xxxx\kill
\FUNCTION $\op{extend-chain}(o, l, j)$ \\
\> \IF $H[l] = \bot$ \THEN \\
\> \> $H[l] \gets \op{hash}(H[l-1] \| o \| l \| j)$ 
         \` // extend by one \\
\> \ELSE \IF $H[l] \neq \op{hash}(H[l-1] \| o \| l \| j)$  \THEN \\
\> \> \RETURN \false
         \` // server replies are inconsistent \\
\> \RETURN \true \\
\\
\FUNCTION $\op{check-view}(\delta, b, a, \psi)$ \\
\> \ASSERT $\op{length}(\delta) \geq 1$ \\
\> \IF $b = c$ \THEN $d \gets c-1$ \ELSE $d \gets c$ \\
\> \FOR $k = 1, \dots, \op{length}(\delta)$ \DO \\
\> \> $l \gets d+k$ \\
\> \> $(o, z, \phi, j) \gets \delta[k]$ \\
\> \> \ASSERT $\op{extend-chain}(o, l, j)$ \\
\> \> \ASSERT $\op{verify}_j(\phi, \str{commit} \| l \| o \| j \| z \| H[l])$ \\
\> \ASSERT $\op{verify}_j(\psi, \str{auth} \| o \| b \| H[b] \| a)$ 
   \` // variables $o$ and $j$ keep their values \\
\> $c \gets d + \op{length}(\delta)$ 
   \` // all operations in $\delta$ have been cleared \\
\\
\FUNCTION $\op{check-pending}(\omega)$ \\
\> \ASSERT $\op{length}(\omega) \geq 1$ \\
\> \FOR $k = 1, \dots, \op{length}(\omega)$ \DO \\
\> \> $l \gets c+k$ \\
\> \> $(o, \tau, j) \gets \omega[k]$ \\
\> \> \ASSERT $\op{extend-chain}(o, l, j)
          \land \op{verify}_j(\tau, \str{invoke} \| o \| j)$ \\
\> \ASSERT $o = u \land j = i$ 
   \` // variables $o$ and $j$ keep their values
\end{tabbing}
}
\caption{ADS integrity protocol (AIP) for client~$C_i$, continued}
\label{alg:client2}
\end{algo}

\begin{algo}
\vbox{
\small
\begin{tabbing}
xxxx\=xxxx\=xxxx\=xxxx\=xxxx\=xxxx\=xxxx\kill
\STATE \\
\> $t \in \N_0$: seqno. of last invoked op., initially 0 \\
\> $b \in \N_0$: seqno. of last applied op., initially 0 \\
\> $I: \N \to \CO \times \{0,1\}^* \times \N$: invoked ops., initially empty \\
\> $O: \N \to \CO \times \{0,1\}^* \times \CZ \times \{0,1\}^* \times \N$: 
   committed ops., initially empty \\
\> $A: \N_0 \to \{0,1\}^* \times \{0,1\}^*$: authenticators, 
   init.~$A[0] = a_0$ \\
\> $s \in \{0,1\}^*$: state of the service, initially $s = s_0$ \\
\\
\UPON receiving message $\msg{invoke}{o, \tau, c}$ from $C_i$ \DO \\
\> $t \gets t + 1$ \\
\> $I[t] \gets (o, \tau, i)$ \\
\> \IF $b = c$ \THEN $\delta \gets \langle O[b] \rangle$ 
      \ELSE $\delta \gets \langle O[c+1], \dots, O[b] \rangle$ \\
\> $\omega \gets \langle I[b+1], I[b+2], \dots, I[t] \rangle$
   \` // all pending operations \\

\> ($\mu, \cdot) \gets \op{separate-pending}(i, \omega) $ \\
\> $(r, \sigma_o) \gets \op{query}_F(s, \mu)$ \\
\> send message \msg{reply}{\delta, b, A[b], \omega, t, r, \sigma_o} 
   to $C_i$ \\
\\
\UPON receiving message $\msg{commit}{o, q, z, \phi}$ from $C_i$ \DO \\
\> $O[q] \gets (o, z, \phi, i)$ \\
\\
\UPON $O[b+1] \neq \bot \land A[b+1] = \bot$ \DO \\
\> $(o, z, \phi, j) \gets O[b+1]$ \\
\> \IF $z = \str{success}$ \THEN \\
\> \> $(r, \sigma_o) \gets \op{query}_F(s, o)$ \\
\> \ELSE \\
\> \> $(r, \sigma_o) \gets (\bot, \bot)$ \\
\> send msg. \msg{update-auth}{o, r, \sigma_o, \phi, b+1, O[b], A[b]} to $C_j$ \\
\\
\UPON receiving message $\msg{commit-auth}{a, \sigma_o, \psi}$ from $C_i$ \DO \\
\> $b \gets b + 1$ \\
\> $A[b] \gets (a, \psi)$ \\
\> $(o, z, \cdot, \cdot) \gets O[b]$ \\
\> \IF $z = \str{success}$ \THEN \\
\> \> $s \gets \op{refresh}_F(s, o, \sigma_o)$
\end{tabbing}
}
\caption{ADS integrity protocol (AIP) for server~$S$}
\label{alg:server}
\end{algo}

This method is common to the server and the clients.  Note that $\mu$
includes the current operation (which appears at the end of~$\omega$)
but not the aborted operations of~$C_i$.  The server finds the status of
a pending-self operation~$o'$ of $C_i$ in $O[b+k]$ (except for $o$
itself, obviously) because $C_i$ has already committed $o'$ prior to
invoking $o$ and because the messages between $C_i$ and $S$ are
FIFO-ordered.  On the other hand, $C_i$ retrieves the status of~$o'$
from~$Z[b+k]$.

Then $S$ computes the response~$r$ and auxiliary data~$\sigma_o$ by
calling $\op{query}_F(s, \mu)$ from the ADS for~$F$; the response
therefore takes into account the state reached after the successful
pending-self operations of $C_i$ but excludes any pending-other
operations present in $\omega$.  However, the client will only
execute~$o$ and output~$r$ when $\gamma$ is compatible with~$o$ and,
therefore, $C_i$ is guaranteed a view in which the operations of $\gamma$
occur after~$o$.  This will ensure \forklin.  The \str{reply} message
to~$C_i$ also includes~$A[b]$ containing the authenticator and its
\str{auth}-signature, for the operation with sequence number~$b$.  The client
passes these to $\op{authexec}_F$ of~$\mu$ for verifying the
correctness of the response.  Furthermore, the \str{reply} message
contains $\delta$, the list of all operations that have been
authenticated since $C_i$'s last operation.  In particular, when $c$
is the sequence number from the \str{invoke} message, $\delta$
contains the operations at sequence numbers $c+1, \dots,b$; when $c = b$,
however, $\delta$ contains still~$O[b]$.

After receiving the \str{reply} message from $S$, the client
(1)~processes and clears the authenticated operations in~$\delta$,
(2)~verifies the pending operations in~$\omega$,
and (3)~verifies that $r$ is the correct response for~$o$.  These steps
are explained next.

For verifying and processing $\delta$ and the last signed authenticator
in~$\alpha$, client~$C_i$ calls a function \op{check-view} and verifies
and/or extends the hash chain for every operation and verifies the
corresponding \str{commit}-signature.  In particular, this ensures for any
operation which has been pending for $C_i$ and must be cleared, that
the \emph{same} operation was also authenticated by its originator.
Finally, $C_i$ also checks the
\str{auth}-signature on the authenticator~$a$, which is contained
in~$\alpha$.  If successful, all operations in $\delta$ are cleared and
$C_i$'s operation counter $c$ is advanced to the position of the last
operation in~$\delta$.  (The check for $b=c$ ensures that $\delta$
contains at least one operation at position~$c$.)

The client continues in \op{check-pending} by verifying that the pending
operations are announced correctly: for every operation in~$\omega$, it
determines the sequence number~$l$, verifies the corresponding
\str{invoke}-signature~$\tau$, and checks the hash chain entry~$H[l]$.
If there is no entry in~$H$ for~$l$, then $C_i$ computes the new entry
from~$o$,~$l$,~$j$, and~$H[l-1]$; otherwise, $C_i$ verifies the that
existing entry matches the expected value.  If this validation succeeds,
it means the operation is consistent with a pending operation sent
previously by~$S$.  After iterating through the pending operations, the
client checks also that the last operation in~$\omega$ is indeed its own
current operation~$o$.

Next, $C_i$ invokes $\op{separate-pending}$ to extract $\mu$ and
$\gamma$ from~$\omega$ (see earlier).  Then, $C_i$ checks whether
$\gamma$ is compatible with~$u$ (the last invoked operation).  If yes,
$C_i$ calls the ADS operation~$\op{authexec}_F(\mu, a, r, \sigma_o)$ for
verifying that applying the operations in~$\mu$ yields $r$ as the
response (recall that $\mu$ includes $o$ at the end).  
The goal of this step is only to check the correctness of the response,
and the authenticator and auxiliary data output by
$\op{authexec}_F$ are ignored.  Finally, $C_i$ commits $o$ by generating
a \str{commit}-signature over~$t$, the sequence number of~$o$, its
status, and its hash chain entry, sends a \str{commit} message
(with~$t$, the operation, and the signature) to~$S$, and outputs the
response~$r$.

\subsubsection{Passive phase}

The server stores the content of all incoming \str{commit} messages in
$O$ and processes them in the order of their sequence numbers, indicated
by~$b$.  When an operation with sequence number~$b+1$ has been committed
but not yet authenticated by the client and applied by~$S$
(i.e., \textbf{upon}~$O[b+1] \neq \bot \land A[b+1] = \bot$), the server
uses $\op{query}_F$ to compute the response~$r$ and to extract the
auxiliary data~$\sigma_o$ from the current state~$s$.  It sends this in
an \str{update-auth} message to $C_i$, also including the operation at
position~$b$ (from $O[b]$) and its authenticator (taken from~$A[b]$),
which have been computed before.  These values allow the client to
verify the authenticity of the response for the operation at
position~$b+1$.

The client~$C_i$ then receives this \str{update-auth} message (for~$o$
and sequence number~$q$), and first validates the message contents.  In
particular, $C_i$ verifies that the authenticator~$a$
is covered by a valid \str{auth}-signature by client~$C_j$ with sequence
number $q-1$, using $C_i$'s hash chain entry~$H[q-1]$.

Next, if $o$ was not aborted, i.e., $Z[q] = \str{success}$, the client
invokes $\op{authexec}_F$ to verify that the auxiliary data and the
response are correct, and to generate new auxiliary data~$s'_o$ and a
new authenticator~$a'$, which vouches for the correctness of the state
updates induced by~$o$.  Otherwise, $C_i$ skips this step, as the
authenticator does not change.  Then $C_i$ issues an
\str{auth}-signature~$\psi'$ and sends it back to $S$ together with $a'$
and $s'_o$ in a \str{commit-auth} message.

As the last step in the passive phase, $S$ increments~$b$, stores the
data in the \str{commit-auth} message at $A[b]$, and if the operation
did not abort, $S$ applies it to~$s$ through~$\op{refresh}_F$.

\subsection{Remarks}
\label{sec:remarks}

As in BST~\cite{wisish09} and in COP~\cite{cacohr14}, operations that do
not interfere with each other may proceed without blocking.
More precisely, if some pending operation is not compatible with the
current operation, the latter is aborted and must be retried later.
Preventing clients from blocking is highly desirable but cannot always
be guaranteed without introducing aborts~\cite{cashsh07}. The potential
for blocking has led other systems, including SPORC~\cite{fzff10} and
FAUST~\cite{cakesh11}, to adopt weaker and less desirable guarantees
than \forklin.

Obviously, it makes no sense for a client to retry its operation while
the non-compatible operation is still pending.  However, the client does
not know when the contending operation commits.  Additional
communication between the server and the clients could be introduced to
signal this.  Alternatively, the client may employ a probabilistic
waiting strategy and retry after a random delay.

In the following we assume that $S$ is correct.  The communication cost
of AIP amounts to the five messages per operation.
Every client eventually learns about all operations of all clients, as
it must clear them and include them in its hash chain.  However, this
occurs only when the client executes an operation (in \str{reply}).  At
all other times between operations, the client may be offline and
inactive.  In a system with $n$ clients that performs $h$ operations in
total, BST~\cite{wisish09} and COP~\cite{cacohr14} require $\Theta(n h)$
messages overall.  AIP reduces this cost to $\Theta(h)$ messages, which
means that each client only processes a small constant number of
messages per operation.

The size of the \str{invoke}, \str{commit}, \str{update-auth}, and
\str{commit-auth} messages does not depend on the number of clients
and on the number operations they execute.  The size of the \str{reply}
message is influenced by the amount of contention, as it contains the
pending operations.  If one client is slow, the pending operations may
grow with the number of further operations executed by other clients.
Note that the oldest pending operation is the one at sequence
number~$b+1$; hence,
all operations ordered afterwards are treated as pending, even when they
already have been committed.  The \str{reply} message can easily be
compressed to constant size, however, by omitting the pending operations
that have already been sent in a previous message to the same client.
See the protocol extensions in Sec.~\ref{sec:extensions} for further
discussion.

The functionality-dependent cost, in terms of communicated state and
auxiliary data, is directly related to the ADS for~$F$.
In practice, hierarchical authenticated search structures, such as hash
trees and authenticated skip lists, permit small authenticators and
auxiliary data~\cite{crowal11}.

\subsection{Correctness}
\label{sec:proof}

We consider three cases: (1) $S$ is correct and the clients execute
operations (1a) sequentially or (1b) concurrently; and (2) $S$ is
malicious.

In case (1a), all operations execute one after each other.  When,
furthermore, the \str{commit-auth} message from a client reaches $S$
before the next operation is invoked, then AIP is similar to
``serialized'' SUNDR~\cite{lkms04} and the ``lock-step protocol'' of
Cachin~\cite{cachin11}.  This means that a client~$C_i$ executing an
operation~$o$ receives a \str{reply} message with all authenticated
operations that exist in the system and only~$o$ as pending operation.
Then $t = b + 1$, and later $C_i$ commits~$o$ and authenticates~$o$
without further operations intermixed at~$S$ nor at any client.
Clearly, this execution is linearizable and satisfies the first
condition of a fork-linearizable Byzantine emulation.

In case (1b), there may exist pending-other operations, but since $C_i$
verifies that its own operations are compatible with them, the response
value is correct.  As $S$ is correct, the views of all clients are
equal, i.e., prefixes of each other, and this ensures linearizability.

For case (2), note that every client~$C_i$ starts to extend its view
from a cleared, authenticated operation and the corresponding signed
authenticator~$a$.
If the pending-other operations in~$\gamma$ are compatible with~$o$ and the
response is valid w.r.t.~$a$ (according to $\op{authexec}_F$), then it is
safe for $C_i$ to output the response and thereby include it into its view.
The malicious~$S$ may order the operations in $\gamma$ differently at other
clients, creating a fork, but they can be omitted from the view of~$C_i$.
The hash chain maintained by $C_i$ contains a condensed representation of
its entire view.  By using its own hash chain entry during the verification
of the \str{commit} and \str{auth} signatures of other clients, $C_i$
ensures that the views of these other clients are equal.  Hence, whenever
an operation~$o$ appears in the views of two clients, also their views are
the same up to~$o$.  This ensures fork-linearizability.

\subsection{Extensions}
\label{sec:extensions}

In order to keep the complexity of the protocol description for AIP at a
comprehensible level, we present important efficiency improvements
informally here.

\paragraph{Eliminating aborted operations.}

The first extension removes aborted operations from being considered as
pending.  Recall that in AIP, an operation~$o$ is included with the
pending operations until the executing client has authenticated~$o$
and the server has processed it during the respective background phase,
even if the operation was aborted.  
This has the drawback that later operations may not be compatible with~$o$
and abort unnecessarily.  However, if $o$ was aborted, this has been
committed by the client, and $S$ has received its \str{commit} message,
then $S$ can include this with the list of pending operations sent for a
later operation~$o'$.  The client that executes~$o'$ will take into account
that~$o$ was aborted and ignore it for determining whether $o'$ is
compatible with the pending operations.  Depending on the workload, this
reduces further aborts.

\paragraph{Batching and delegating operation authentication.}

Recall that the clients authenticate operations in the order of the
server-assigned sequence numbers.  Some client~$C_\op{slow}$ may fall
behind with this step, and when the other clients proceed faster and
execute more operations, the number of pending operations grows
continuously.  This creates much more work for the faster clients for
processing the \str{reply} message and slows them down.

However, since all clients trust each other, one can modify AIP such
that another client~$C_\op{fast}$ may step in for $C_\op{slow}$, handle
the \str{update-auth} message, and sign the authenticator for the
operation of~$C_\op{slow}$.  Only small changes to the data structures
are needed to accommodate this change.  Ideally $C_\op{fast}$ has more
processing power or is closer to~$S$ on the network than~$C_\op{slow}$;
this choice should be determined heuristically based on the measured
performance or network delays.

Extending the above idea, $S$ may actually batch all non-authenticated
operations when an operation from~$C_i$ commits at sequence number~$q$.
Hence, $S$ sends the \str{update-auth} messages for all operations between
$b$ and~$q$ to~$C_i$ and delegates the step of authenticating them
to~$C_i$.  This works because a client can authenticate two consecutive
operations without going back to~$S$.  The server may then record the
\str{commit-auth} responses from the fastest client and inform the others
that their help with authenticating operations is no longer needed.

\paragraph{Passive phase only for update operations.}

Recall that the protocol executes a functionality~$F$ whose operations can
be separated into query and update operations ($\CQ$ and $\CU$,
respectively).  Queries do not change the state; as is easy to see, they
are compatible with \emph{every} subsequent operation.  But the passive
phase of AIP is only needed for creating a new authenticator, after the
state has changed.  Therefore we can eliminate the passive phase for all
query operations; this considerably improves the efficiency of the protocol
for read-intensive applications.

In particular, for every operation $o \in \CQ$, the passive phase is
skipped and the verification operations are adjusted accordingly.  When a
client $C_i$ has committed a query operation, the server immediately
``applies'' it and does not send an \str{update-auth} message later.  For
implementing this, the sever has to maintain another variable~$d$ with the
sequence number of the operation that most recently modified the state. The
\str{reply} and \str{update-auth} messages now contain the operation $O[d]$
that allows clients to verify the corresponding authenticator~$A[d]$.

\paragraph{Tolerate client crashes.}

Crashes of clients are not included in the formal model here, but must be
considered in practice.  In order to tolerate crashes, we reuse the second
extension above.  Since a client $C_\op{fail}$ may crash before it finishes
both protocol phases, the server will never receive the missing message(s)
and the operation is never authenticated.  Going beyond our formal model,
suppose that another client $C_i$ has determined that $C_\op{fail}$ has
crashed.  Then $C_i$ may take over, abort the hanging operations, and
commit and authenticated in place of the failed client.  It is important
that $C_\op{fail}$ must not execute any operations again later, hence the
failure detection should be reliable; if $C_\op{fail}$ may join again
later, it must be given a new identity.  Group management protocols for
adding and removing clients dynamically have been discussed in the context
of existing systems, such as VENUS~\cite{scckms10} and SPORC~\cite{fzff10}.

\section{Verification of integrity and consistency of cloud object
  storage~(\NAME)}
\label{sec:frieda}

We are now ready to introduce our main contribution, the protocol for
\emph{verifying the integrity and consistency of cloud object storage},
abbreviated \emph{\NAME}.  It leverages AIP from the previous section and
provides a fork-linearizable Byzantine emulation for a practical
object-store service, in a manner that is transparent to the storage
provider.  We first define the operations of the cloud storage service and
outline the architecture of \NAME.  Next we instantiate AIP for verifying
the integrity of a simple object store and show how \NAME extends this to
practical cloud storage.

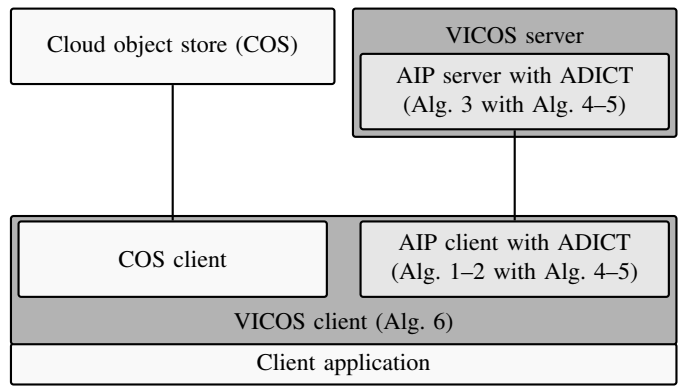
\begin{figure}[ht]
    \begin{center}

    \begin{tikzpicture}[
            thick,
            font=\footnotesize,
            every node/.style={
                draw,
                align=center,
                text width=4cm,
                rounded corners=1pt,
                minimum height=1cm,
            }
        ]

        \node[text width=8.5cm, minimum height=0.5cm, fill=gray!5] (client) {Client application};

        \node[text width=8.5cm, fill=gray!60, minimum height=1.7cm, above=-1pt of client] (frieda) {};
        \node[draw=none, text width=8.5cm, minimum height=0.5cm, above=-1pt of client] {\NAME client (Alg.~\ref{alg:frieda})};

        \node[fill=gray!5] (store) [above=3.1cm of frieda.west, anchor=west] {Cloud object store (COS)} ;
        \node[fill=gray!5,text width=3.8cm, xshift=0.1cm] (cos-client) [below=1.1cm of frieda.north west, anchor=south west] {COS client};

        \node[fill=gray!20, text width=3.8cm, xshift=-0.1cm] (aip-client) [below=1.1cm of frieda.north east, anchor=south east] {AIP client with ADICT\\(Alg.~\ref{alg:client}--\ref{alg:client2} with Alg.~\ref{alg:ad_aip1}--\ref{alg:ad_aip2})};

        \node[fill=gray!60, minimum height=1.7cm] (vicos-server) [above=2.75cm of frieda.east, anchor=east] {\NAME server};    
        \node[fill=gray!20, text width=3.8cm, xshift=-0.1cm, yshift=0cm] (aip-server) [above=2.5cm of frieda.east, anchor=east] {AIP server with ADICT\\(Alg.~\ref{alg:server} with Alg.~\ref{alg:ad_aip1}--\ref{alg:ad_aip2})};
        \node[draw=none, minimum height=0.5cm] (vicos-server) [above=3.25cm of frieda.east, anchor=east] {\NAME server};
        
        \draw[thick] (store.south) to [] (store.south |- cos-client.north);
        \draw[thick] (aip-server.south) to [] (aip-server.south |- aip-client.north);
    \end{tikzpicture}

    \end{center}     

    \caption{Architecture of \NAME: the two untrusted components of the cloud 
      service are shown at the top, the trusted client is at the bottom.}
    \label{fig:frieda-algorithm-overview}
\end{figure}

More precisely, \NAME consists of the following components (see
Fig.~\ref{fig:frieda-algorithm-overview}):
\begin{enumerate}
\item A \emph{cloud object store (COS)} service with a key-value store
  interface, as offered by commercial providers.  It maintains the object
  data stored by the clients using~\NAME.
\item An \emph{AIP client} and an \emph{AIP server}, which implement the
  protocol from the previous section for the functionality of an
  \emph{authenticated dictionary (ADICT)} and authenticate the objects at
  the cloud object store.  The AIP server runs remotely as a cloud service
  accessed by the AIP client.  This is abbreviated as \emph{AIP with
    ADICT}.
\item The \emph{\NAME client} exposes a cloud object store interface to the
  client application and transparently performs integrity and consistency
  verification.  During each operation, the client consults the cloud
  object store for the object data itself and the AIP server for
  integrity-specific metadata.  In particular, AIP server running ADICT
  stores a cryptographic hash of every object.
\end{enumerate}
Note that the cloud object store as well as the AIP server are in the
untrusted domain; they may, in fact, collude together against the
clients.

\subsection{Cloud object store (COS)}
\label{sec:kvs}

The \emph{cloud object store} is modeled by \emph{key-value store (KVS)}
and provides a ``simple'' storage service to multiple clients.  It stores a
practically unbounded number of objects in a flat namespace, where each
\emph{object} is an arbitrary sequence of bytes (or a ``blob,'' a binary
large object), identified by a unique \emph{name} or \emph{key}.  We assume
that clients may only read and write entire objects, i.e., it is not
possible to read from or write into the middle of an object, as in a file
system.

Our formal notion of a KVS internally maintains a map~$M$ that stores
the values in~$\CV$ under their respective keys taken from a
universe~$\CK$.  It provides four operations:
\begin{enumerate}\shortlist
\item $\op{kvs-put}(k, v)$: Stores a value~$v \in \CV$ under key~$k \in
  \CK$, that is, $M[k] \gets v$.
\item $\op{kvs-get}(k)$: Returns the value stored under key~$k \in
  \CK$, that is, $M[k]$.
\item $\op{kvs-del}(k)$: Deletes the value stored under key~$k \in
  \CK$, that is, $M[k] \gets \bot$.
\item $\op{kvs-list}()$: Returns a list of all keys for which
  a value is stored, that is, the list $\langle k \in \CK | M[k]
  \neq \bot \rangle$.
\end{enumerate}
This API forms the core of many real-world cloud storage services, such
as Amazon S3 or OpenStack Swift.  Typically there is a bound on the
length of the keys, such as a few hundred bytes, but the stored values
can be much larger and practically unbounded (on the order of several 
Gigabytes).  For simplicity, we assume
that the object store provides atomic semantics during concurrent
access, being aware that cloud storage systems may only be eventually
consistent~\cite{baigho13} due to network partitions.

Many practical cloud object stores support a single-level hierarchical
name space, formed by \emph{containers} or \emph{buckets}.  We abstract
this separation into the keys here; however, a production-grade system
would introduce this separation again by applying the design of \NAME
for every container.

\subsection{Authenticated dictionary implementation (ADICT)}
\label{sec:aip_ad}

\begin{algo}[!tbh]
\vbox{
\small
\begin{tabbing}
xxxx\=xxxx\=xxxx\=xxxx\=xxxx\=xxxx\=xxxx\kill
\STATE \\
\> $D: \CK \to \{0,1\}^*$: authenticated dictionary, initially empty \\
\> $T$: hash tree over~$D$, initially empty \\
\\
\FUNCTION $\op{query}_{\op{ADICT}}((D, T), o)$ \\
\> \IF $o = \op{adict-put}(k, v) \lor o = \op{adict-del}(k)$ \THEN \\
\> \> $r \gets \bot$ \\
\> \> $s_o \gets \text{sibling nodes on path (*) from $k$ to root in $T$}$ \\
\> \ELSE \IF $o = \op{adict-get}(k)$ \THEN \\
\> \> $r \gets D[k]$ \\
\> \> $s_o \gets \text{sibling nodes on path (*) from $k$ to root in $T$}$ \\
\> \ELSE \ \ // $o = \op{adict-list}()$ \\
\> \> $r \gets \langle k \in \CK | D[k] \neq \bot \rangle$ \\
\> \> $s_o \gets T$ \\
\> \RETURN $(r, s_o)$ \\
\\
\FUNCTION $\op{authexec}_{\op{ADICT}}(o, a, r, s_o)$ \\
\> \IF $o = \op{adict-put}(k, v) \lor o = \op{adict-get}(k) \lor o = \op{adict-del}(k)$
\THEN \\
\> \> \IF $s_o$ is not a valid path (*) from $k$ to tree root~$a$ \THEN \\
\> \> \> \RETURN $(\cdot, \cdot, \false)$ \\
\> \IF $o = \op{adict-put}(k, v)$ \THEN \\
\> \> insert leaf node $k$ with value $v$ in the tree \\
\> \> $s'_o \gets \text{updated path from $k$ to tree root}$ \\
\> \> $a' \gets \text{updated hash-tree root}$ \\
\> \ELSE \IF $o = \op{adict-get}(k)$ \THEN \\
\> \> \IF path (*) not consistent with node~$k$ holding $r$ \THEN \\
\> \> \> \RETURN $(\cdot, \cdot, \false)$ \\
\> \> $s'_o \gets \bot$ \\
\> \> $a' \gets a$ \\
\> \ELSE \IF $o = \op{adict-del}(k)$ \THEN \\
\> \> delete leaf node $k$ from the tree \\
\> \> $s'_o \gets \text{updated paths from siblings of $k$ to tree root}$ \\
\> \> $a' \gets \text{updated hash-tree root}$ \\
\> \ELSE \ \ // $o = \op{adict-list}()$ \\
\> \> \IF $r$ is not list of keys in leaves of tree with root~$a$ \THEN \\
\> \> \> \RETURN $(\cdot, \cdot, \false)$ \\
\> \> $s'_o \gets \bot$ \\
\> \> $a' \gets a$ \\
\> \RETURN $(a', s'_o, \true)$
\end{tabbing}
}
\caption{Authenticated dictionary implementation (ADICT) for AIP, Part~1}
\label{alg:ad_aip1}
\end{algo}

\begin{algo}[!tbh]
\vbox{
\small
\begin{tabbing}
xxxx\=xxxx\=xxxx\=xxxx\=xxxx\=xxxx\=xxxx\kill
\FUNCTION $\op{refresh}_{\op{ADICT}}((D, T), o, s_o)$ \\
\> \IF $o = \op{adict-put}(k, v)$ \THEN \\
\> \> $D[k] \gets v$ \\
\> \> update path in $T$ from $k$ to root, as taken from $s_o$ \\
\> \ELSE \IF $o = \op{adict-del}(k)$ \THEN \\
\> \> $D[k] \gets \bot$ \\
\> \> update path in $T$ from $k$ to root, as taken from $s_o$ \\
\> \RETURN $(D, T)$ \\
\\
\FUNCTION $\op{compatible}_{\op{ADICT}}(\mu, u)$ \\
\> \FOR $o \in \mu$ \DO \\
\> \> \IF $\neg \op{compatible}_{ADICT}(u, o)$ \THEN 
\` // See Table~\ref{tab:com-op} for the $\op{compatible}_{ADICT}()$ relation
  on operations\\
\> \> \> \RETURN \false \\
\> \RETURN \true
\end{tabbing}
}
\caption{Authenticated dictionary implementation (ADICT) for AIP, Part~2}
\label{alg:ad_aip2}
\end{algo}

\NAME instantiates AIP with the functionality of a KVS that stores only
\emph{short values}.  In order to distinguish it from the cloud object
store, we refer to it as \emph{authenticated dictionary}, denoted by
\op{ADICT}, with operations \op{adict-put}, \op{adict-get},
\op{adict-del}, and \op{adict-list}.

The implementation of ADICT uses the well-known approach of building a hash
tree over its entries~\cite{Blum:1994,naonis00,crowal11}; see
Alg.~\ref{alg:ad_aip1}--\ref{alg:ad_aip2} for the details of how ADICT is
implemented within AIP.  The AIP server stores the values in a
map~$D$ and maintains a hash tree~$T$, constructed over the list of
key-value pairs stored in the map, according to a fixed sort order on the
keys.  That is, every leaf node of the hash tree is computed by hashing the
node key, its value, and the key of the successor leaf node together.  The
next node has to be included in order to authenticate the \emph{absence} of
a key in response to a query for a non-existing key~(e.g.,
\cite{naonis00}).  The root of the hash tree serves as the authenticator
for ADICT.

For the \op{adict-put}, \op{adict-get}, and \op{adict-del} operations, the
server extracts those paths from $T$ that are necessary to verify the
correctness of the retrieved value and places them in~$s_o$.  For
\op{adict-put} and \op{adict-del} operations, $\op{query}_{\op{ADICT}}$
also places these paths into~$s_o$ because the client needs them to
construct the updated root hash.  For \op{adict-list}, the complete hash
tree is included in $s_o$.
The asterisks~(*) in Alg.~\ref{alg:ad_aip1}--\ref{alg:ad_aip2} denote some
additional data and steps necessary to verify the predecessor or
successor leaves for authenticating an absent key (details of this are
omitted here and can be found in the literature~\cite{crowal11}).

The $\op{compatible}_{\op{ADICT}}(\mu, u)$ function of
Alg.~\ref{alg:ad_aip1}--\ref{alg:ad_aip2} defines the compatibility
relation among the operations of the authenticated dictionary; \NAME
supports the same KVS interface and inherits this notion of compatibility
for the cloud-storage operations.  For more general services like
databases, one would invoke a transaction manager here.  For ADICT, the
compatibility between a first (pending) operation~$u$ and a second
(current) operation~$o$ is given by Table~\ref{tab:com-op}.  For instance,
\op{adict-put} followed by \op{adict-get} for the same key or followed by
\op{adict-list} are not compatible, whereas two \op{adict-list} and
\op{adict-get} operations are always compatible.

The advantage of considering operation compatibility over commutativity (as
used in ACOP~\cite{cacohr14}, for instance) becomes apparent here: only 8
pairs among the 49 cases shown are not compatible, whereas 22 among 49
cases do not commute and would be aborted with commutativity.

\begin{table}
    \centering
    \begin{tabular}{r@{\hskip 2em}ccccccc}
        \toprule
        & \rotatebox{90}{$\op{adict-put}(x,\cdot)$} & \rotatebox{90}{$\op{adict-put}(y,\cdot)$} & \rotatebox{90}{$\op{adict-get}(x)$} & \rotatebox{90}{$\op{adict-get}(y)$} & \rotatebox{90}{$\op{adict-del}(x)$} & \rotatebox{90}{$\op{adict-del}(y)$} & \rotatebox{90}{$\op{adict-list}$} \\

        \hline
        $\op{adict-put}(x,\cdot)$   & $\surd$ & $\surd$ & ---     & $\surd$ & $\surd$ & $\surd$ & --- \\
        $\op{adict-put}(y,\cdot)$   & $\surd$ & $\surd$ & $\surd$ & ---     & $\surd$ & $\surd$ & --- \\
        $\op{adict-get}(x)$ & $\surd$ & $\surd$ & $\surd$ & $\surd$ & $\surd$ & $\surd$ & $\surd$ \\
        $\op{adict-get}(y)$ & $\surd$ & $\surd$ & $\surd$ & $\surd$ & $\surd$ & $\surd$ & $\surd$ \\
        $\op{adict-del}(x)$   & $\surd$ & $\surd$ & ---     & $\surd$ & $\surd$ & $\surd$ & --- \\
        $\op{adict-del}(y)$   & $\surd$ & $\surd$ & $\surd$ & ---     & $\surd$ & $\surd$ & --- \\
        $\op{adict-list()}$     & $\surd$ & $\surd$ & $\surd$ & $\surd$ & $\surd$ & $\surd$ & $\surd$ \\
        \bottomrule
    \end{tabular}
  \caption{The $\op{compatible}_{ADICT}(\cdot, \cdot)$ relation for
    ADICT and the KVS interface, where $x, y \in \CK$ denote distinct keys}
  \label{tab:com-op}
\end{table}

\clearpage

\subsection{\NAME client implementation}
\label{sec:prototype}

\NAME emulates the key-value store API of a cloud object store (COS) to
the client and transparently adds integrity and consistency
verification.
As with AIP, consistency or data integrity violations committed by the
server are detected through \ASSERT; any failing assertion triggers an
alarm.
It must be followed by a recovery action whose details go beyond the
scope of this paper.  Analogously to AIP, \NAME may return \str{abort};
this means that the operation was not executed and the client should
retry it.

\begin{algo}
\vbox{
\small
\begin{tabbing}
xxxx\=xxxx\=xxxx\=xxxx\=xxxx\=xxxx\=xxxx\kill
\FUNCTION $\op{\MakeLowercase{\NAME}-put}(k, v)$ \\
\> $x \gets \text{a random nonce}$ \\
\> $\op{cos-put}(k \| x, v)$ \\
\> $h \gets \op{hash}(v)$ \\
\> $r \gets  \op{aip-invoke}(\op{adict-put}(k, \langle x, h \rangle))$ \\
\> \IF $r = \str{abort}$ \THEN
   \` // concurrent incompatible operation \\
\> \> $\op{cos-del}(k \| x)$ \\
\> \RETURN $r$ \\
\\
\FUNCTION $\op{\MakeLowercase{\NAME}-get}(k)$ \\
\> $r \gets \op{aip-invoke}(\op{adict-get}(k))$ \\
\> \IF $r = \str{abort}$ \THEN 
   \` // concurrent incompatible operation \\
\> \> \RETURN $\str{abort}$ \\
\> $\langle x, h \rangle \gets r$ \\
\> $v \gets \op{cos-get}(k \| x)$ \\
\> \ASSERT $\op{hash}(v) = h$ \\
\> \RETURN $v$ \\
\\
\FUNCTION $\op{\MakeLowercase{\NAME}-del}(k)$ \\
\> $r \gets \op{aip-invoke}(\op{adict-del}(k))$ \\
\> \IF $r = \str{abort}$ \THEN
   \` // concurrent incompatible operation \\
\> \> \RETURN $\str{abort}$ \\
\> \op{cos-del}($k \| *$) 
   \` // deletes all keys with prefix $k$ \\
\> \RETURN $r$ \\
\\
\FUNCTION $\op{\MakeLowercase{\NAME}-list}()$ \\
\> $r \gets \op{aip-invoke}(\op{adict-list}())$ \\
\> \IF $r = \str{abort}$ \THEN
   \` // concurrent incompatible operation \\
\> \> \RETURN \str{abort} \\
\> \RETURN $r$
\end{tabbing}
}
\caption{Implementation of \NAME at the client.}
\label{alg:frieda}
\end{algo}

Algorithm~\ref{alg:frieda} presents the pseudo~code of the \NAME client. Basically, it
protects every object in the COS by storing its cryptographic hash in the authenticated
dictionary (ADICT).  Operations on the object store trigger corresponding
operations on COS and on ADICT, as provided by AIP for consistency
enforcement.

In order to prevent race conditions, \NAME does not store the hash of an
object under the object's key in COS directly, but \emph{translates} every
object key to a unique key for COS.  Otherwise, two concurrent
operations accessing the same object might interfere with each other and
leave the system in an inconsistent state.
More precisely, in a $\op{\MakeLowercase{\NAME}-put}(k, v)$ operation, the
client chooses a nonce~$x$ (a value guaranteed to be unique in the system,
such as a random string) and stores $v$ in COS using $\op{cos-put}(k \| x,
v)$ with the translated key $k \| x$. Furthermore, it computes $h \gets
\op{hash}(v)$ and stores $\langle x, h \rangle$ in AD using \op{adict-put}
at key~$k$.  The \op{cos-put} and \op{cos-get} operations actually stream
long values.  When \op{adict-put} aborts due to concurrent operations, the
client deletes $v$ again from COS using $\op{cos-del}(k \| x)$.

For a $\op{\MakeLowercase{\NAME}-get}(k)$ operation, the client first
calls $\op{adict-get}(k)$ and retrieves $\langle x, h \rangle$.  Unless
this operation aborts, the client translates the key and calls
$\op{cos-get}(k \| x)$ to retrieve the value~$v$.  After $v$ has been
read (or streamed), the client compares its hash value to~$h$,
asserts that they match, and then outputs~$v$.

Without key translation, two concurrent $\op{\MakeLowercase{\NAME}-put}$
operations~$o_1$ and $o_2$ writing different values to the same key~$k$
might both succeed with $\op{cos-put}(k, v_1)$ and $\op{cos-put}(k,
v_2)$, respectively, but the \op{adict-put} for~$o_2$ might abort due to
another concurrent operation.  Then COS might store $v_2$ but ADICT
stores the hash of $v_1$ and readers would observe a false integrity
violation.
Thanks to key translation, no versioning conflicts arise in the COS.
Atomicity for multiple operations on the same object key follows from the
properties of AIP with the ADICT implementation.
The $\op{\MakeLowercase{\NAME}-del}(k)$ and
$\op{\MakeLowercase{\NAME}-list}()$ operations proceed analogously; but
the latter does not access~COS.

\subsection{Correctness}
\label{sec:frieda_correctness}

It is easy to see that the implementation of \NAME satisfies the two
properties of a fork-linearizable Byzantine emulation.  First, when $S$ is
correct, then the clients proceed with their operations and all
verification steps succeed.  Hence, \NAME produces a linearizable
execution.  The linearization order is established by AIP running ADICT.
Furthermore, when the clients execute sequentially, then by the
corresponding property of AIP, no client ever receives \str{abort} from
ADICT.

Second, consider the case of a malicious server controlling COS and the AIP
server together.  AIP ensures that the operations on ADICT (\op{adict-put},
\op{adict-get}, etc.) are fork-linearizable according to
Sec.~\ref{sec:frieda}.  The implementation of ADICT follows the known
approach of memory checking with hash trees~\cite{Blum:1994,naonis00} and
therefore authenticates the object hash values that \NAME writes to ADICT.
According to the properties of the hash function, the object data is
uniquely represented by its hash value.  Since \NAME ensures the object
data written to COS or returned to the client corresponds to the hash value
stored in ADICT, it follows that all operations of \NAME are also
fork-linearizable.

\section{Prototype}
\label{sec:implementation}

We have implemented a prototype of \NAME in Java; it consists of a
client-side library (``\NAME client'') and the server code (``\NAME
server'').  The system can be integrated with applications that require
cryptographic integrity and consistency guarantees for data in untrusted
cloud storage services.  It is available as
open-source on GitHub~(\url{https://github.com/ibm-research/vicos}).

\subsection{\NAME implementation}

The client-side library uses the \emph{BlobStore} interface of Apache
\emph{jclouds} (Version~2.0.0~-Snapshot,
\url{https://jclouds.apache.org/}) for connecting to different cloud object
stores.  The server runs as a standalone web service, communicating with
the client-side library using the \emph{Akka} framework~(Version 2.4.4,
\url{http://www.akka.io}).

Akka is an event-driven framework which supports the actor
model~\cite{Hewitt:1973:UMA:1624775.1624804}.  It fits perfectly into the
system model of \NAME and allows for rapid development.  This simplifies
the actual protocol implementation because of the high level of
abstraction, especially relating to network operations and concurrency.

The client library as well as the server code are implemented as
actors within the framework.  Actors are independent units which can only
communicate by exchanging messages.  Every actor has a mailbox that buffers
all incoming messages.  By default messages are processed in FIFO order by
the actor.  This allows the server protocol implementation to process all
incoming messages sequentially and execute each protocol step atomically,
that is, mutually exclusive with respect to all others.  The implementation
of \NAME therefore closely follows the high-level description according to
Sec.~\ref{sec:generic-protocol}.  Note that this clearly limits the systems
performance by not utilizing modern multi-core architecture, on the other
hand.

We developed the \NAME client library so that it may easily be integrated
into existing applications to provide integrity protection.  It uses the
modular approach of AIP instantiated with an authenticated data structure
(ADS).  A developer only needs to implement the desired functionality, by
defining the state (e.g., the internal KVS data structure), a set of
operations, and their compatibility relation.  For that reason, we have
defined two interfaces: \emph{state} and \emph{operation processor}.
Operations are described using Google's Protocol
Buffers~(\url{https://developers.google.com/protocol-buffers/}) executed by
the operation processor.  This modular concept allows us to reuse and
extend the core implementation of the protocol.

The \NAME prototype provides a simple KVS functionality by implementing
these interfaces.  More precisely, the KVS state is a map supporting
\str{get}, \str{put}, \str{del} and \str{list} operations.  The KVS operation
processor provides the implementations of \op{query}, \op{authexec},
\op{refresh}, and \op{compatible} as described in
Alg.~\ref{alg:ad_aip1}--\ref{alg:ad_aip2}.  The client and the server
protocol each contain an instance of the KVS operation processor.  The client
exposes a simple KVS interface, adding Java Exceptions for signaling
integrity and consistency violations.  Furthermore, the client library is
completely asynchronous and supports processing the AIP passive phase
in the background without blocking the client process.

The cryptographic signatures can be implemented in multiple ways.
According to the security assumptions of the model, all clients trust each
other, the server alone may act maliciously, and only clients issue digital
signatures.  Therefore, one can also adopt a simplified trust model with
``signatures'' provided by a message-authentication code (MAC).  For many
applications, where strong mutual trust exists among the clients, MACs
suffice and will result in faster execution.  On the other hand, this
simplification renders the system more fragile and exposes it easily to
attacks by clients.

In particular, \NAME uses HMAC-SHA1 with 128-bit keys provided by the Java
Crypto\-graphy Extension as the default signature implementation.  The code
also supports RSA and DSA signatures with 2048-bit keys.  A user can choose
between these signature implementations via a configuration file.  This
approach also allows developers to change to a different signature
implementation, or even implement their own according to new requirements.
Particularly, this might be useful for porting the code to other platforms
such as mobile devices, with less computation power.

The core implementation of \NAME consists of $\appr$3400~sloc, the
server part is $\appr$400~sloc more, whereas the client part including
the integration with the evaluation platform (see below) takes
$\appr$800~sloc extra.

\subsection{Practical issues and optimizations}
\label{sec:practialissues}

While developing \NAME and experimenting with it, we obtained experience
with Akka and gained insight into the protocol's operation itself.  This
has led to further optimizations described here.

\paragraph{Bounded pending list.}

An issue that we discovered in Akka while implementing \NAME is Akka's
default maximum message size of only~\SI{128}{\kilo\byte}.  In particular,
this becomes a problem in \NAME when \str{reply} messages include a large
partial state or a very long list of pending operations.  By configuring Akka with larger
message sizes (we used a maximum size of~\SI{128}{\mega\byte}), the direct
limitation disappears.

However, we experienced that large messages impact the overall performance
negatively.  When the number of pending operations increases, the resulting
very large messages slow down the operations of \NAME.
Therefore, we implemented a way to bound the length of the
pending-operations list, that is, we introduced a maximum number of pending
operations as a configurable value and modified the protocol.  When this
maximum is reached, the server buffers all new incoming requests
(\str{invoke} messages) until enough other operations have completed
and the number of pending operations goes  below the limit.
We tested with different maximum
sizes for the pending list from 32 up to 1024  operations, and chose
a limit of 128 for the evaluations.

A more robust solution of that issue would be to signal the clients to wait
before sending more requests, instead of just buffering them at the server.
Although limiting the number of pending operations under high server load
increases the latency of client requests, it also increases the overall
performance and stability of the system.  In summary we found that the
benefits of this optimization outweigh its drawbacks.

\paragraph{Message delivery order.}  

During development we experienced slow performance caused by the FIFO order
in which the protocol actors processed the arriving messages. Therefore, we
implemented priority mailboxes for the server and the client actor and
defined a priority rule to prefer \str{commit}, \str{update-auth},
\str{commit-auth} messages over \str{invoke} and \str{reply} messages.
This has the immediate benefit that the server processes \str{update-auth}
messages and thereby completes the \str{passive phase} of already
authenticated operations \emph{before} it starts working on new
\str{invoke} messages.  This preference shortens the list of pending
operations directly.  Certainly, this may increase the response time for
new operations again, but eventually, it prevents more operations from
aborting due to conflicts under high load.

\section{Evaluation}
\label{sec:evaluation}

This section reports on performance measurements with the \NAME prototype.
They study the general overhead of integrity protection, the scalability of
the protocol, and the effect of (in-)compatible operations.

\subsection{Experimental setup}

The experiments use cloud servers and an OpenStack Swift-based object
storage service (\url{http://swift.openstack.org}) of a major cloud
provider with about two dozen data centers world-wide (Softlayer~--- an
IBM Company, \url{http://www.softlayer.com/object-storage}).

The \NAME server runs on a dedicated ``baremetal'' cloud server with a
3.5GHz Intel Xeon-Haswell (E3-1270-V3-Quadcore) CPU, \SI{8}{GB} DDR3 RAM,
and a \SI{1}{Gbps} network connection.
The clients run on six baremetal servers in total, each server with 2x~2GHz
Intel Xeon-SandyBridge (E5-2620-HexCore) CPUs, \SI{16}{GB} DD3 RAM, and a
\SI{1}{Gbps} network connection.  All clients are hosted in the same data
center. All machines run Ubuntu~14.04-64 Linux and Oracle~Java (JRE~8,
build 1.8.0\_77-b03).

To simulate a realistic environment, we conduct experiments in two settings
as shows in Table~\ref{tab:eva:settings}.  A~\emph{datacenter} setting,
with all components in the same data center (``Amsterdam''), and a
\emph{wide-area} setting, where the VICOS server and COS are located
together in one data center (``Milan''), and the clients at a remote site
(``Amsterdam'').

\begin{table}[ht!]
    \centering

    \begin{tabular}{@{}l llll@{}}
        \toprule
        Setting & Clients & \NAME server & Cloud object storage & Latency [ms] \\

        \midrule
        Datacenter & Amsterdam & Amsterdam & Amsterdam & $<1$\\
        Wide-area & Amsterdam & Milan & Milan & $\sim10$\\
        \bottomrule
    \end{tabular}

    \caption{Evaluation setting}
    \label{tab:eva:settings}
\end{table}

The datacenter setting establishes a best-case baseline due to the very low
network latencies (\SI{< 1}{\ms}).  This deployment is not very realistic
in terms of the security model because the clients and the storage service
are co-located.

The wide-area setting exhibits a moderate network latency~(round-trip delay
time of \SI{\appr 20}{\ms}) between the two data centers and models the
typical case of geographically distributed clients accessing a cloud
service with its point-of-access on the same continent but in different
countries.

\begin{figure}[ht]
  \centering
  \resizebox{14cm}{!}{\input{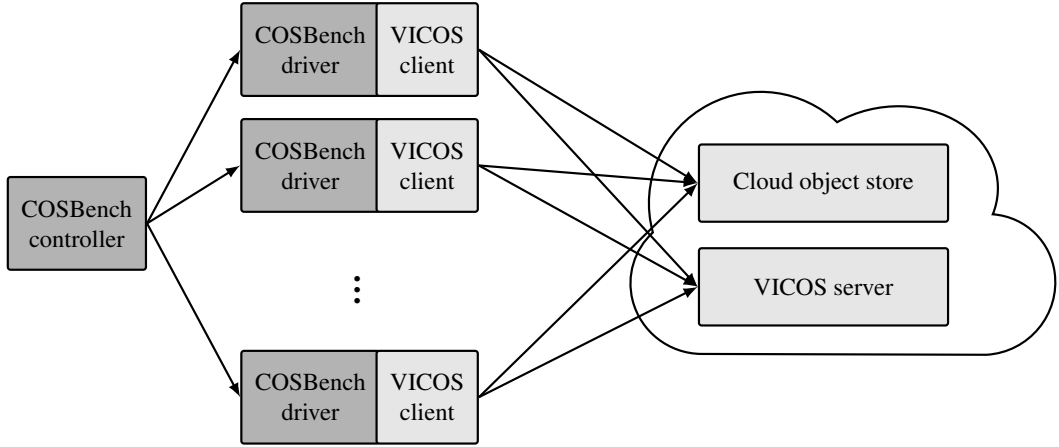}\unskip}
  \caption{The experimental setup, with one COSBench controller and many
    COSBench drivers, accessing the cloud storage service through the
    \NAME client.}
  \label{fig:cosbench}
\end{figure}

The evaluation is driven by COSBench (Version 0.4.2 -- \url{https://github.com/intel-cloud/cosbench}), an extensible tool for
benchmarking cloud object stores. We have created an adapter to drive
\NAME from COSBench, as shown in Fig.~\ref{fig:cosbench}.
COSBench uses a distributed architecture, consisting of multiple
\emph{drivers}, which generate the workload and simulate many clients
invoking concurrent operations on a cloud object store, and one
\emph{controller}, which controls the drivers, selects the workload
parameters, collects results, and outputs aggregated statistics.  In
particular, the COSBench setup for \NAME reports the \emph{operation
  latency}, defined as the time that an operation takes from invocation
to completion, and
the aggregated \emph{throughput}, defined as the data rate between all
clients and the cloud storage service.

COSBench invokes ``read'' and ``write'' operations, implemented by
\op{\MakeLowercase{\NAME}-get} and \op{\MakeLowercase{\NAME}-put},
respectively.  Every reported data point involves read and write operations
taken over a period of at least \SI{30}{\s} after a \SI{30}{\s} warm-up.
In the experiments two configurations are measured:
\begin{enumerate}\shortlist
\item The \emph{native} object storage service as a baseline, with direct
  unprotected access from COSBench to cloud storage, but accessing the
  cloud storage through the \emph{jclouds} interface; and
\item \emph{\NAME}, running all operations from COSBench through
  \emph{jclouds} and the verification the protocol.
\end{enumerate}

\subsection{Results}

\subsubsection{Cryptography microbenchmark}

In a first experiment we study how different signature implementation
affect the computation and network overhead of~\NAME.
We implemented digital signatures using RSA and DSA with~\SI{2048}{\bit}
keys, and additionally HMAC-SHA1 with~\SI{128}{\bit} keys.  The
cryptographic algorithms are provided by the SunJCE version~1.8 provider.

\begin{figure}[ht!]
    \centering
    \includegraphics[width=14cm]{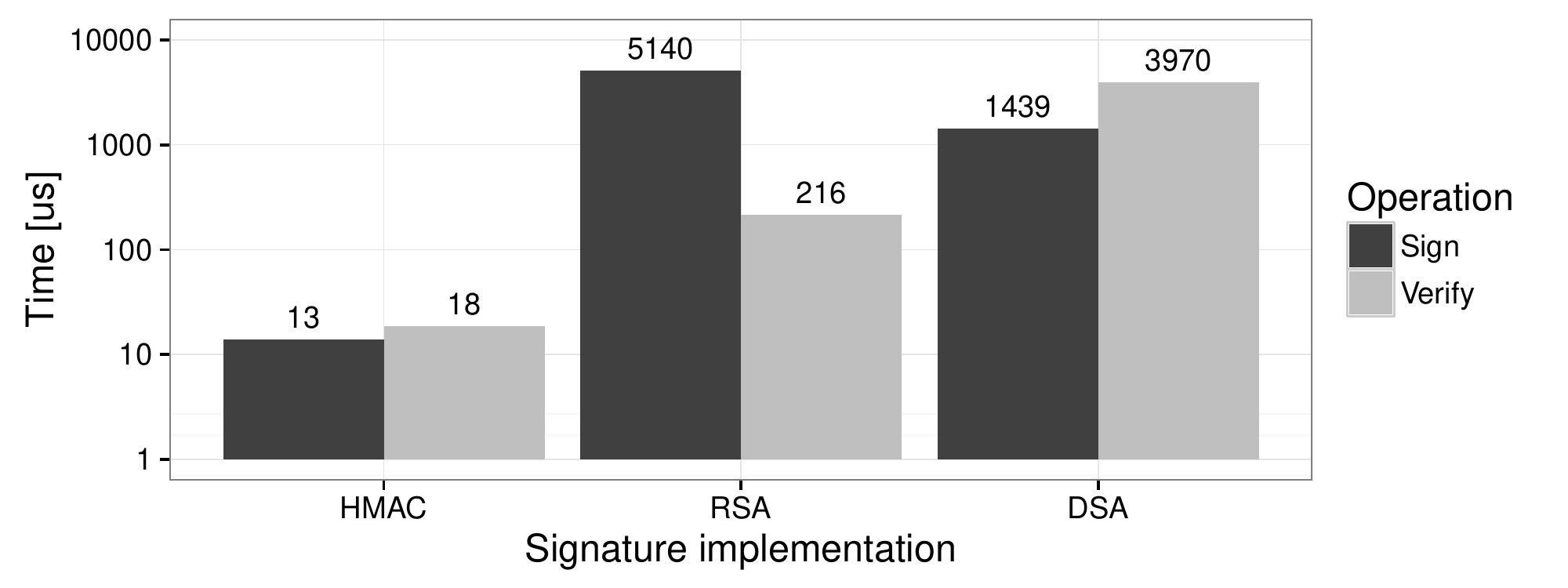}

    \caption{
    The average time for digital-signature operations (HMAC, RSA, and DSA);
    note that the y-axis uses log-scale.}
    \label{fig:eval:micro-crypto}
\end{figure}

We measured the time it takes on a client to sign and verify an
invoke message using the three signature implementations.
Figure~\ref{fig:eval:micro-crypto} shows that the RSA signing takes
around~\SI{5}{\ms}, while the verification takes around~\SI{220}{\us}.  DSA
takes around~\SI{4}{\ms} for signing and around~\SI{1.5}{\ms} for
verification, whereas HMAC signing and verification only takes less
than~\SI{20}{\us}.
Additionally, the resulting signature sizes have a direct effect on the
message sizes and the network load. RSA signatures are~\SI{256}{byte},
while DSA signatures are only~\SI{40}{byte}.  HMAC reduces the signature
size to~\SI{20}{byte}.

We use HMAC-SHA1 as the default implementation of signatures in
the remainder of the evaluation.  Its operations are much faster than for
RSA and DSA signatures, reducing the computation overhead at the client.
Moreover, the smaller signature size of HMAC-SHA1 also reduces the network
overhead of~\NAME.

\subsubsection{Object size}
\label{sec:evalblobsize}

\begin{figure}[htb!]
    \centering
    \includegraphics[width=14cm]{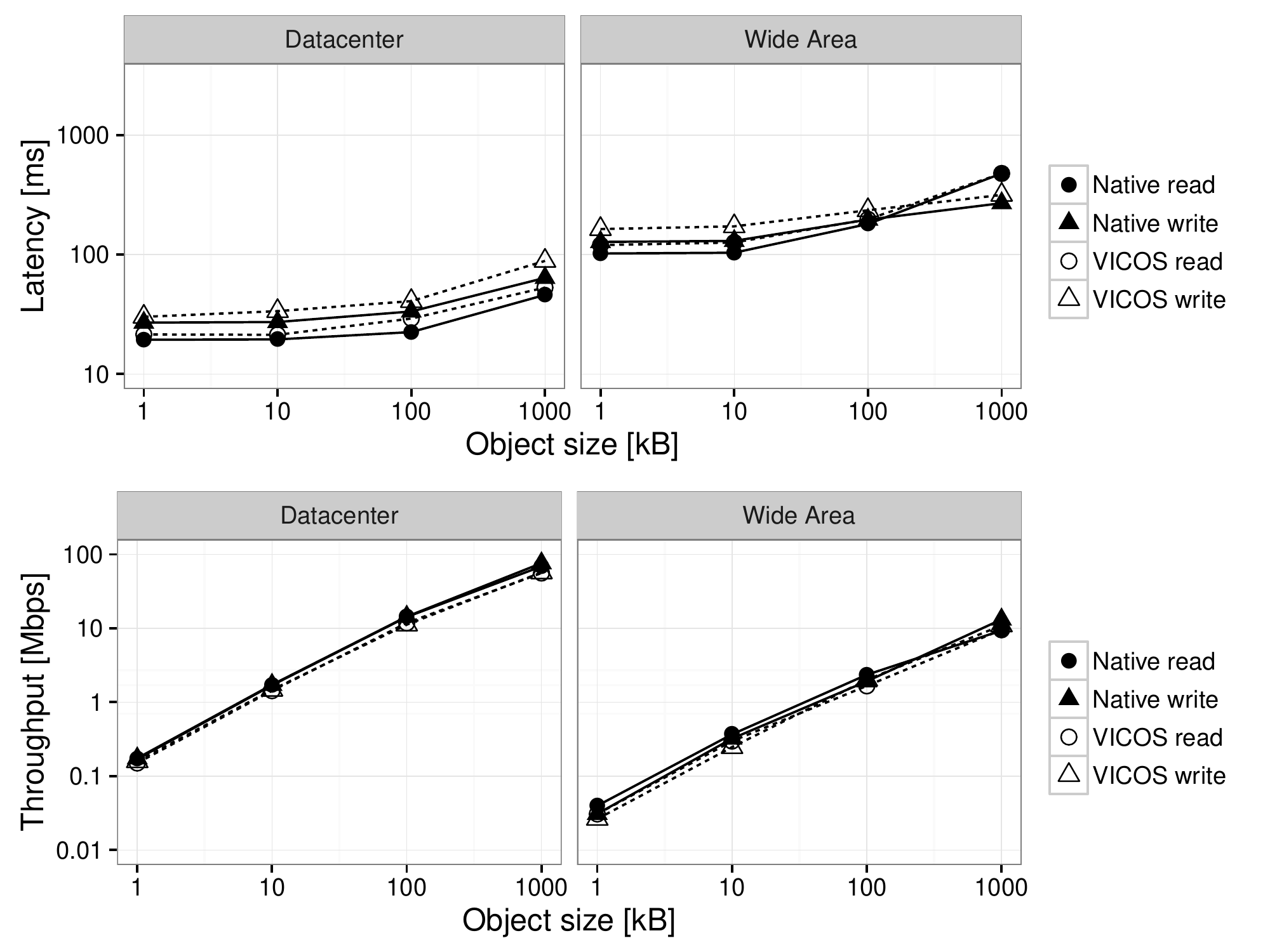}

    \caption{
    The effect of different object sizes:
    Latency and throughput of read and write operations with one client.}
    \label{fig:eval:objectsize}
\end{figure}

In this experiment we study how the object size affects the latency and the
throughput of \NAME.  We define a workload with a single client executing
read and write operations for objects of size
\SIlist[exponent-to-prefix]{1;10;100;1e3}{\kilo\byte}.

Fig.~\ref{fig:eval:objectsize} shows that the latency and throughput of
\NAME behave very similar to the native system.  As expected, we observe
that \NAME introduces an overhead that incurs a small cost compared to
unprotected access to storage.  In particular, for the datacenter setting,
\NAME increases latency by an average of 16.2\% for \emph{read}, and 24.0\%
for \emph{write}; it decreases throughput by an average of 15.8\% for
\emph{read}, and 17.7\% for \emph{write}.  We also expect the overhead to
decrease with bigger objects.  However, we could not find this effect in
the datacenter setting: here the overhead remains practically constant,
from the small to the large objects.  In the wide-area setting, the
overhead is approximately the same for the smaller objects
(\SI{1}{\kilo\byte} and \SI{10}{\kilo\byte}), but it indeed decreases
as the object size grows and disappears at the largest object size
(\SI{1000}{\kilo\byte}).

Interestingly, in the wide-area setting, the relative performance in terms
of throughput is reversed between \emph{read} and \emph{write} for
\SI{1000}{\kilo\byte} objects.  Whereas \emph{read} has lower latency and
achieves better throughput than \emph{write} in all other experiments, this
relation is reversed in the right-most data points of
Fig.~\ref{fig:eval:objectsize}.  This may be caused by caching data on the
cloud object store, which improves read performance for smaller objects,
but disappears when larger objects are stored and accessed less frequently
than in the datacenter setting.

\subsubsection{Number of clients}
\label{sec:evalclients}

\begin{figure}[!ht]
    \centering
    \includegraphics[width=14cm]{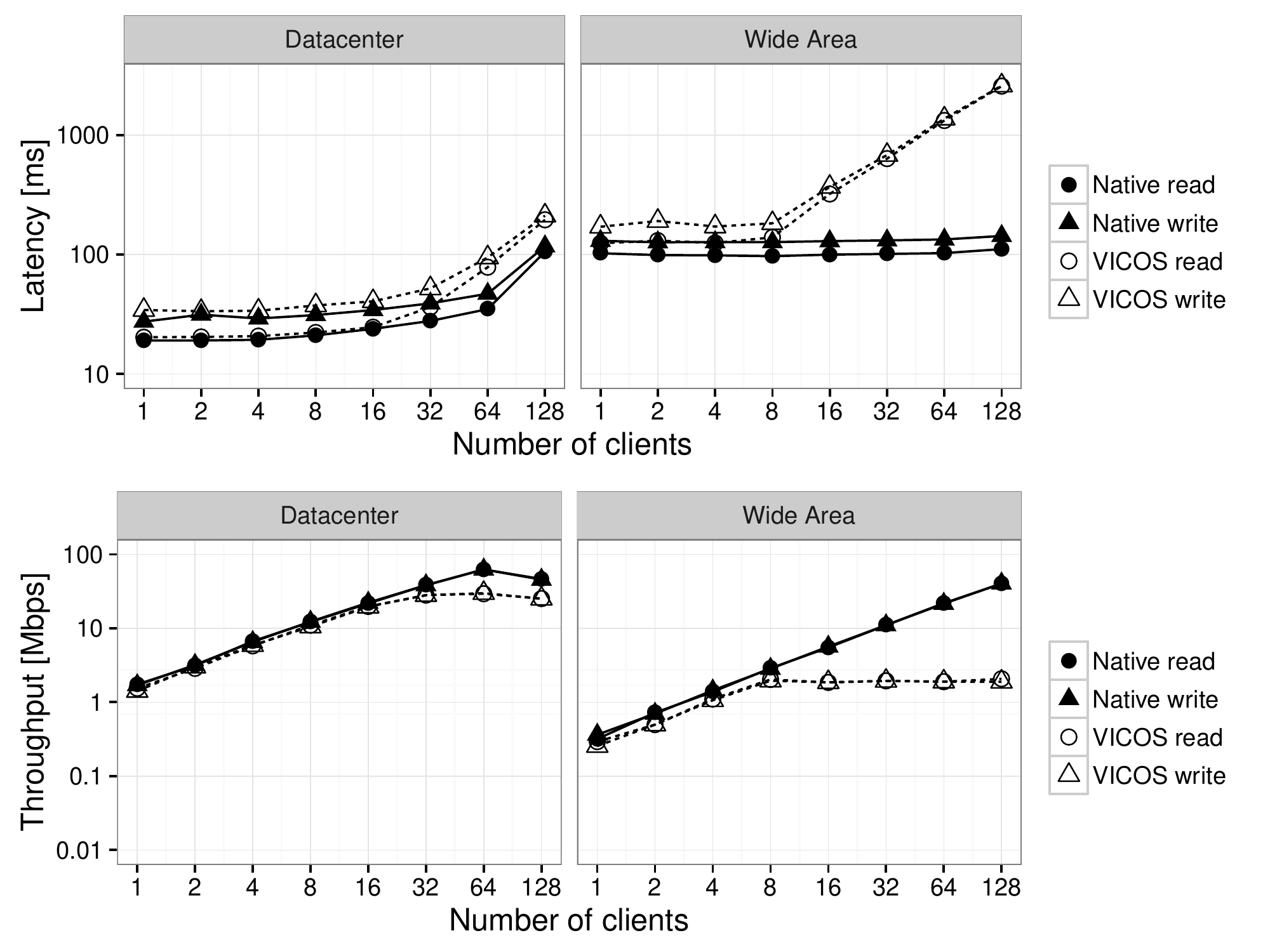}

    \caption{
    Scalability with the number of clients:
    Latency and throughput of read and write operations with
    \SI{10}{\kilo\byte} objects.}
    \label{fig:eval:clients}
\end{figure}

We also study the scalability of \NAME by increasing the number of
clients.  The workload uses up to~\num{128}~clients (spread uniformly
over the six COSBench drivers), and \num{64} objects with a fixed size
of~\SI{10}{\kilo\byte}. One half of the objects are designated for read
operations and the other half for write operations, respectively.  This
division prevents concurrency conflicts among the client operations.
As mentioned in Sec.~\ref{sec:practialissues}, we restrict the size
of the pending list in the protocol to~\num{128} operations.
We do not use a large number of clients because of the underlying
assumption that clients trust each other, which might not be realistic in
much larger groups.

As Fig.~\ref{fig:eval:clients} shows, the native system throughput
scales linearly until the network is saturated with~\num{64} clients in
the data center setting. \NAME follows the same behavior but reaches
saturation already with~\num{32} clients.
In contrast, in the wide-area setting, \NAME becomes saturated with \num{8}
clients, from where throughput remains almost constant and latency grows.
No saturation is evident with the native configuration and up to 128
clients.
 
The reason for the slower operation in the wide-area experiment is that all
requests of the clients are handled by the \NAME server sequentially and
thus it becomes a bottleneck of the system.
Due to the higher latency in the wide-area setting, operations remain
longer in the pending queue.  This means more work for the clients and
the server.  Since the active and passive protocol phases
are asynchronous, clients may invoke the next operation already before they
have completed a previous operation.  Hence, they reach limit of the
bounded pending queue and the throughput of \NAME remains limited at the
rate imposed by the server's operation.

\subsubsection{Concurrent operations}
\label{sec:evalconflict}

\begin{figure}[t]
    \centering

    \includegraphics[width=14cm]{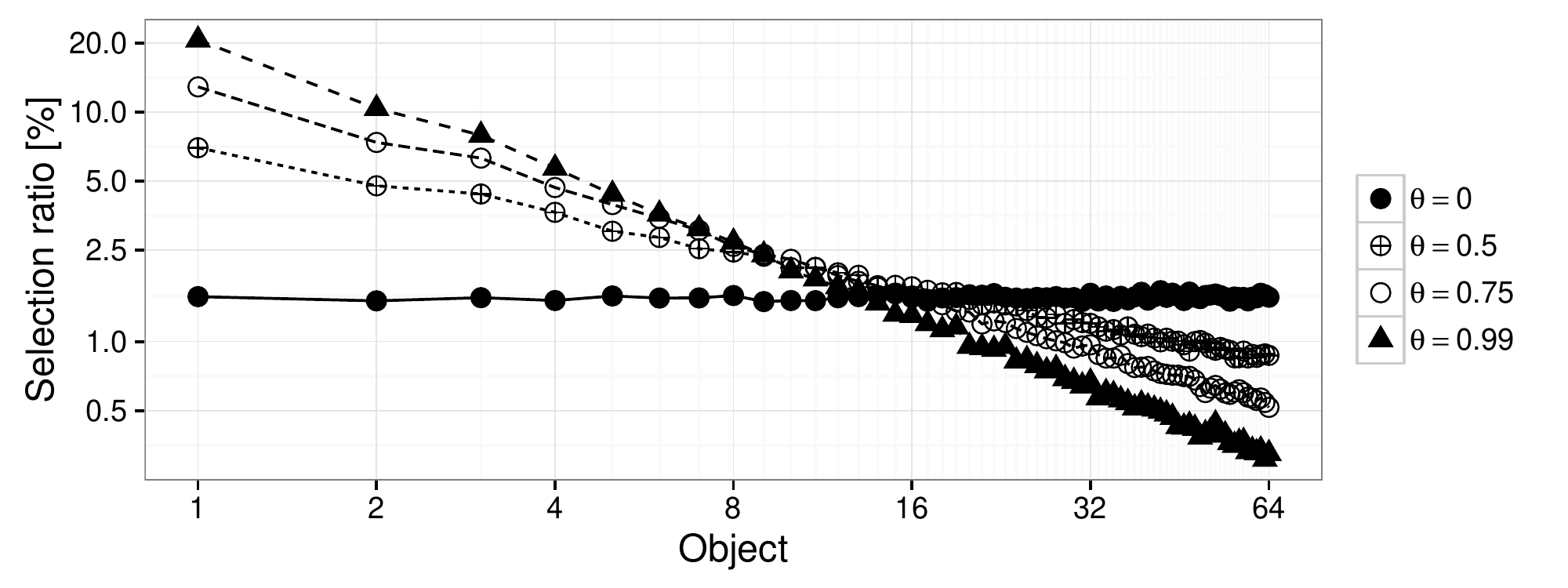}

    \caption{
    The effect of different values for the Zipf distribution
    factor~$\theta$ of the COSBench object selector: Selection rate for
    each object with 10000 selections, 64 objects, and varying
    $\theta$.}
    \label{fig:eval:zipf-dist}
\end{figure}

\begin{figure}
    \centering

    \includegraphics[width=14cm]{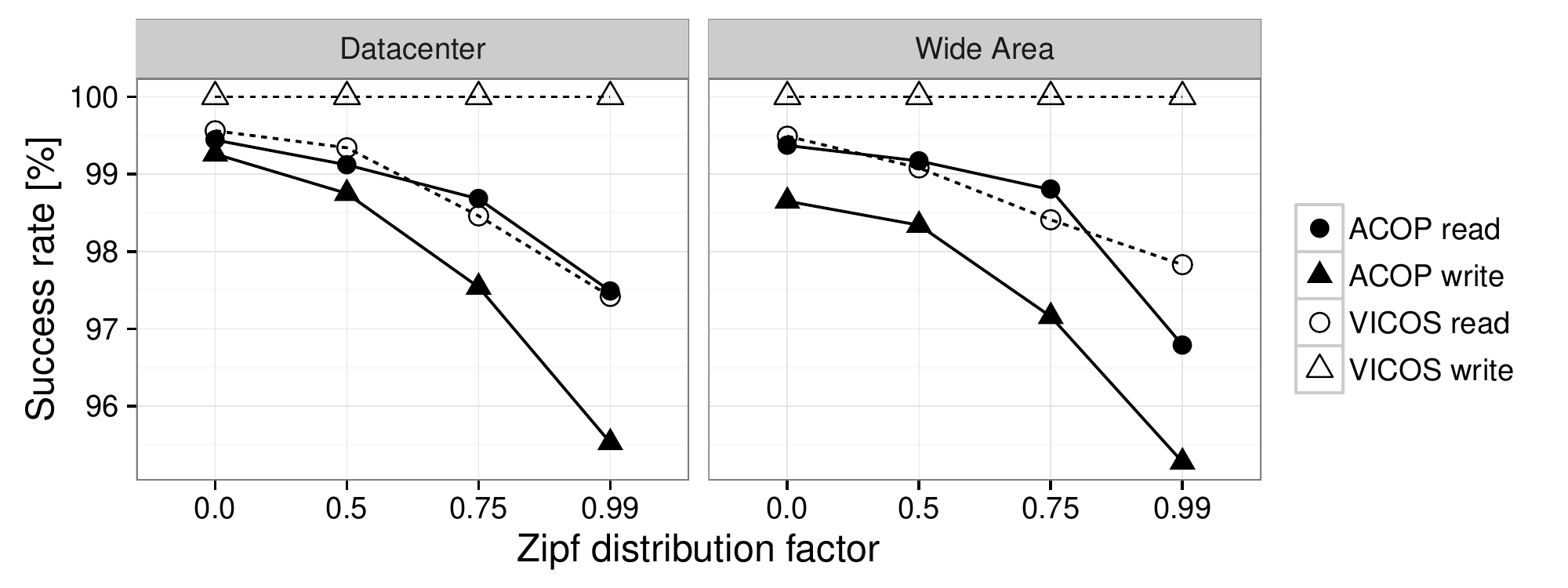}

    \caption{
    The effect of conflicting concurrent operations: Success rate of
    read and write operations with \SI{10}{\kilo\byte} objects, sixteen
    clients, and varying Zipf distribution factors~$\theta$.}
    \label{fig:eval:zipf}
\end{figure}

Finally, we investigate the effect of conflicting concurrent operations.
\NAME aborts an operation if it is not compatible with one of the pending
operations (according to Sec.~\ref{sec:defcompatible}). In that case the
client has to retry later.
Protocols like BST~\cite{wisish09} and ACOP~\cite{cacohr14} are more
cautious and abort as soon as two operations do \emph{not commute}, which
occurs more often.  Hence, we define ACOP as our baseline for this
experiment.
The implementation throws an abort exception when a conflict occurs; that
causes COSBench to report the operation as failed and to continue
immediately with the next operation.  At the end of each experiment COSBench
reports the overall operation success rate.  We expect a higher success
rate for~\NAME compared to ACOP as already discussed in
Sec.~\ref{sec:aip_ad}.

To evaluate this behavior we created a workload with sixteen clients each
invoking read and write operations over~\num{64} objects with a fixed size
of~\SI{10}{\kilo\byte}.  Object accesses are chosen according to ``Zipf's
law'', which approximates many types of data series found in natural and
social structures.  The Zipf distribution is based on a ranking of the
elements in a universe and postulates that the frequency of any element is
inversely proportional to its rank in the frequency table.  Thus, the most
``popular'' object will occur approximately twice as often as the second
most popular one, three times as often as the third most popular and so on.
Zipf distributions are often observed when users access
websites~\cite{adamic2002zipf}.  Figure~\ref{fig:eval:zipf-dist} shows the
object access rate using four different values for the Zipf
factor~$\theta$.  For $\theta=0.99$ the access rate for the first object
(with the biggest contention) is about 20\%, and the least often accessed
object is selected only with probability about 0.3\%.  With $\theta=0$ the
Zipf distribution corresponds to uniformly random access over the objects.

Since COSBench does not support Zipf distributions by default, we
implemented the algorithm of Gray et al.~\cite{gsebw94} for generating a
Zipf-like access distribution, as also used in
YCSB~(\url{https://github.com/brianfrankcooper/YCSB}).
With this workload we cause operations to conflict by progressively
increasing $\theta \in \{0, 0.5, 0.75, 0.99\}$.  The higher the Zipf
factor~$\theta$, the more clients concurrently access the same objects and
invoke non-compatible operations, causing aborts.

Figure~\ref{fig:eval:zipf} shows the operation success rates for \NAME and
for ACOP, using the four different Zipf factors.  Recall from
Sec.~\ref{sec:defcompatible} and from Table~\ref{tab:com-op} that a
\op{put} operation in the KVS interface is always compatible with every
preceding operation and never aborts.  Therefore, the \emph{write}
operations show 100\% success rate for \NAME.  For the commuting operations
in ACOP, on the other hand, \emph{writes} are progressively more often
aborted with increasing~$\theta$.  The behavior of \emph{reads} is similar
for \NAME and ACOP because preceding \emph{writes} cause aborts equally
often.

The advantage of~\NAME over protocols considering only operation
commutativity becomes evident here, in that \emph{write} operations always
succeed, and overall the abort rate is significantly reduced.

\subsubsection{Summary}

The performance evaluation shows that \NAME achieves its goal of adding
consistency and integrity protection while remaining almost transparent to
clients using cloud object stores.  The cost added over the raw performance
is most visible in the datacenter setting, which is not a realistic
deployment for the intended applications.  Still this overhead remains
limited to about 20\% for accesses with high throughput
(Sec.~\ref{sec:evalblobsize}).  With many clients performing operations
concurrently, the extra cost may become noticeable
(Sec.~\ref{sec:evalclients}), and further work is needed for decreasing
this.  It should be added that the \NAME prototype is currently a proof of
concept and not product-level code.

\section{Related Work}
\label{sec:related_work}

Many previous systems providing data integrity rely on trusted
components.
Distributed file systems with cryptographic protection provide
stronger notions of integrity and consistency than given by \NAME; 
there are many examples for this, from early research
prototypes like FARSITE~\cite{abcccd02} or SiRiUS~\cite{gsmb03} to 
production file-systems today (e.g., IBM Spectrum Scale,
\url{http://www-03.ibm.com/systems/storage/spectrum/scale/}).
However, they rely on trusted directory services for freshness.  Such a trusted
coordinator is often missing or considered to be impractical.
Iris~\cite{Stefanov:2012:ISC:2420950.2420985} relies on a trusted
gateway appliance, which mediates all requests between the clients and
the untrusted cloud storage.
Several recent systems ensure data integrity with the help of trusted
hardware, such as CATS~\cite{Yumerefendi:2007:SAN:1288783.1288786}, which
offers accountability based on an immutable public publishing medium, or
A2M~\cite{Chun:2007:AAM:1323293.1294280}, which assumes an append-only
memory.  They all require some form of global synchronization, usually done
by the trusted component, for critical metadata to ensure linearizability.
In the absence of such communication, as assumed here, they cannot protect
consistency and prevent replay attacks.

In CloudProof~\cite{plmwz11}, an object-storage protection system with
accountable and proof-based data integrity and consistency support, clients
may verify the freshness of returned objects with the help of the data
owner.  Its auditing operation works in epochs and verifies operations on
one object only with a certain probability and only at the end of an epoch.
Moreover, the clients need to communicate directly with the owner of an
object for establishing integrity and consistency.

Cryptographic integrity guarantees are of increasing interest for many
diverse domains: Verena~\cite{kfpc16}, for example, is a recent enhancement
for web applications that involve database queries and updates by multiple
clients.  It targets a patient database holding diagnostic data and
treatment information.  In contrast to \NAME, however, it relies on a
trusted server that supplies hash values of data objects to clients during
every operation.

The remainder of this section discusses related work without trusted 
components for synchronization.
With only one client, the classic solution for memory checking by Blum
et al.~\cite{Blum:1994} provides data integrity through a hash tree
and by storing its root at the client.  Many systems have exemplified
this approach for remote file systems and for cloud storage (e.g.,
Athos~\cite{Goodrich:2008:AEA:1432478.1432486}).

With \emph{authenticated data structures}~\cite{naonis00,mndgks04}, the
single-writer, multi-reader model of remote storage can be authenticated,
assuming there is a trusted and timely way to distribute authenticators
from the writer to all readers.  In practice, this approach is often taken
for software distribution, where new releases are posted to a repository
and authenticated by broadcasting hash values of the packages over a
mailing list.  AIP as introduced in Sec.~\ref{sec:generic-protocol}
represents one way to generalize ADS for multiple writers.

In the multi-client model, Mazi{\`e}res et al.~\cite{mazsha02,lkms04} have
introduced the notion of \forklin and implemented SUNDR, the first system
to guarantee fork-linearizable views to all clients.  It detects integrity
and consistency violations among all clients that become aware of each
other's operations.  The SUNDR system uses messages of size~$\Omega(n^2)$
for $n$ clients~\cite{cashsh07}, which might be expensive.  The SUNDR
prototype~\cite{lkms04} description also claims to handle multiple files
and directory trees; however, the protocol description and guarantees are
stated informally only, so that it remains unclear whether it achieves
\forklin under all circumstances.

As mentioned in Sec.~\ref{sec:introduction}, several systems have expanded
the guarantees of \forklin to different applications~\cite{fzff10} and
improved the general efficiency of protocols for achieving
it~\cite{cashsh07}.  Others have explored aborting operations~\cite{mdss09}
or introduced weak \forklin in order to avoid blocking operations.  In
particular, SPORC~\cite{fzff10}, FAUST~\cite{cakesh11}, and
Venus~\cite{scckms10} sacrifice full linearizability to avoid aborts and
blocking, respectively, and achieve weak \forklin instead. The latter is a
relaxation of \forklin in which the most recent operation of a client may
violate atomicity.

The SPORC system~\cite{fzff10} is a groupware collaboration service whose
operations may conflict with each other, but can be made to commute by
applying a specific technique called ``operational transformations.''
Through this mechanism, different execution orders still converge to the
same state; still SPORC achieves only weak \forklin.

Furthermore, \NAME also reduces the communication overhead compared to past
systems considerably, since SUNDR, FAUST, and Venus all use messages of
size $\Theta(n)$ or more with $n$~clients, whereas the message size in
\NAME does not depend on~$n$.

The BST protocol~\cite{wisish09} supports an encrypted remote database
hosted by an untrusted server that is accessed by multiple clients.  Its
consistency checking algorithm allows some commuting client operations to
proceed concurrently; COP and ACOP~\cite{cacohr14} extend BST and also
guarantee \forklin for arbitrary services run by a Byzantine server, going
beyond data storage services, and support wait-freedom for commuting operations.
\NAME builds directly on COP, but improves the efficiency by avoiding the
local state copies at clients and by reducing the computation and
communication overhead.  The main advantage is that clients can remain
offline between executing operations without stalling the protocol.

\section{Conclusion}
\label{sec:conclusion}

This paper has presented \NAME, a complete system for protecting the
integrity and consistency of data outsourced to untrusted commodity cloud
object stores.  It shows, for the first time, how to realize multi-client
integrity protection for generic functions with an authenticated data
structure (ADS).  Its two-phase protocol structure reduces the
communication overhead compared to previous algorithms.

\NAME works with commodity cloud storage services and ensures the best
possible consistency notion of \forklin.  It supports wait-free client
operations and does not require any additional trusted components.

There are several challenges that this paper not address, which remain
open for future work.  An interesting question, for instance, is how to
recover from an integrity violation.  Since we assume only a single
untrusted server and that client data resides at the cloud storage
service, orthogonal techniques are needed for resilience of the data itself.

Another interesting challenge would be to consider malicious clients, as
one further step towards a more realistic system.  For small groups of
clients our system model makes sense, but for groups with hundreds of
clients it seems difficult to maintain this assumption.  The situation is
especially interesting when a client colludes with the malicious server.

Finally, the approach of AIP also be applied to services beyond cloud
storage; for example, cloud and NoSQL databases, interactions in a
social network, or certificate and key management services.

\section*{Acknowledgments}

We thank R\"udiger Kapitza, Dieter Sommer and S\"oren Bleikertz for helpful
discussions and the anonymous reviewers of SYSTOR 2015 for valuable
comments.

This work has been supported in part by the European Commission through the
Horizon 2020 Framework Programme (H2020-ICT-2014-1) under grant agreements
number 644371~WITDOM and 644579~ESCUDO-CLOUD and in part by the Swiss State
Secretariat for Education, Research and Innovation (SERI) under contracts number
15.0098 and 15.0087.

\bibliographystyle{IEEEtranS}
\bibliography{vicos}

\end{document}